\documentclass{jfm}
\usepackage[utf8x]{inputenc}
\usepackage[T1]{fontenc}

\usepackage[dvipsnames]{xcolor}

\definecolor{aqua}{rgb}{0.0, 1.0, 1.0}
\usepackage[normalem]{ulem}
\usepackage{censor}
\newlength\nextcharwidth
\makeatletter
\renewcommand\@cenword[1]{%
  \setlength{\nextcharwidth}{\widthof{#1}}%
  \censorrule{\nextcharwidth}%
  \kern -\nextcharwidth%
  #1}
\makeatother

\usepackage{paralist} 

\usepackage{float}
\usepackage{placeins}

\usepackage{tabu}
\usepackage{booktabs,dcolumn}
\usepackage{multirow,bigdelim}

\usepackage{graphicx}
\usepackage{epstopdf, epsfig}

\usepackage{latexsym}
\usepackage{amssymb}
\usepackage{amsmath}
\usepackage{mathtools}
\usepackage{mathrsfs}
\usepackage{times}
\usepackage{pifont}
\usepackage{bbding}
\usepackage[outline]{contour}
\newcommand{\graylozenge}{\contour{black}{$\color[rgb]{1,1,1}{\blacklozenge} \,$}}

\newcommand{\graytriangleright}{\contour{black}{$\color[rgb]{0.5,0.5,0.5}{\blacktriangleright} \,$}}
\newcommand{\graytriangleleft}{\contour{black}{$\color[rgb]{0.5,0.5,0.5}{\blacktriangleleft} \,$}}
\newcommand{\graystar}{\contour{black}{$\color[rgb]{0.5,0.5,0.5}\bigstar \,$}}
\newcommand{\grayfourstar}{\contour{black}{{\color[rgb]{0.5,0.5,0.5}\FourStar}}}
\newcommand\solidrule[1][1cm]{\hspace{2mm}\rule[0.5ex]{#1}{1pt}\hspace{2mm}}
\newcommand\dashedrule{\mbox{%
  \hspace{2mm}\rule[0.75ex]{2mm}{1pt}\hspace{1mm}\rule[0.75ex]{2mm}{1pt}\hspace{1mm}\rule[0.75ex]{2mm}{1pt}\hspace{2mm}}}
\newcommand\dottedrule{\mbox{%
  \hspace{2mm}\rule[0.75ex]{0.5mm}{1pt}\hspace{0.2mm}\rule[0.75ex]{0.5mm}{1pt}\hspace{0.2mm}\rule[0.75ex]{0.5mm}{1pt}\hspace{0.2mm}\rule[0.75ex]{0.5mm}{1pt}\hspace{0.2mm}\rule[0.75ex]{0.5mm}{1pt}\hspace{0.2mm}\rule[0.75ex]{0.5mm}{1pt}\hspace{0.2mm}\rule[0.75ex]{0.5mm}{1pt}\hspace{0.2mm}\rule[0.75ex]{0.5mm}{1pt}\hspace{0.2mm}\rule[0.75ex]{0.5mm}{1pt}\hspace{2mm}}}

\usepackage{siunitx} 
\DeclareSIUnit[]{\pixel}{px}

\usepackage{natbib}
\usepackage[obeyspaces]{url}

\shorttitle{Boundary layer footprint on separated flows}
\shortauthor{F. Stella, N. Mazellier and A. Kourta}

\title{On scaling of mass entrainment in separated shear layers: the footprint of the incoming boundary layer}

\author{Francesco Stella,
  Nicolas Mazellier
  \corresp{\email{nicolas.mazellier@univ-orleans.fr}}
 \and Azeddine Kourta}

\affiliation{Univ. Orléans, INSA-CVL, PRISME, EA 4229, F45072, Orléans, France}

\begin{document}

\maketitle

\begin{abstract}
We experimentally investigate the effects on scaling of separating/reattaching flows of the ratio $\delta_e/h$, where $\delta_e$ is the thickness at separation of the incoming boundary layer and $h$ the characteristic cross-stream scale of the flow. In the present study, we propose an original approach based on mean mass entrainment, which is the driving mechanism accounting for the growth of the separated shear layer. The focus is on mass transfer at the Turbulent/Non-Turbulent Interface (TNTI). In particular, the scaling of the TNTI, which is well documented in turbulent boundary layers, is used to trace changes in the scaling properties of the flow. To emphasise the influence of the incoming boundary layer, two geometrically similar, descending ramps with sizeably different heights $h$ but fundamentally similar values of $\delta_e$ are compared. The distribution in space of the TNTI highlights a sizeable footprint of the incoming boundary layer on the separated flow, the scaling of which results of the competition between $h$ and $\delta_e$. On the basis of a simple mass budget within the neighbourhood of separation, we propose to model this competition by introducing the scaling factor $C_{h,\delta} = 1 + \delta_e/h$. With this model, we demonstrate that the relationship between shear layer growth and mass entrainment rates established for free shear layers (i.e. $\delta_e = 0$) might be extended to flows where $\delta_e/h > 0$. Since many control systems rely on mass entrainment to modify separation properties, our findings suggest that the parameter $\delta_e/h$ needs to be taken into account when choosing the most relevant strategies for controlling or predicting separating/reattaching flows.
\end{abstract}

\section{Introduction}\label{sec:intro}
Geometry-triggered separating/reattaching flows are common in industrial applications. Their undesirable effects, including degraded aerodynamic performances, increased vibrations and noise, motivate the great efforts which have been dedicated to investigating these flows (see for example the review by \citet{nadge14}).
On the backward-facing step (BFS) and other prototypical salient-edge bluff bodies, separation is fixed by geometry. Downstream of separation, the flow develops into a wide shear layer (\citet{simpson1989}), which grows until it impinges the wall at the reattachment point, several step heights downstream of the BFS. 
The region of the flow lying between the wall and the shear layer is called recirculation region, due to the reversed direction of the main velocity component. The local depression induced by the recirculation region is at the core of many negative effects of flow separation, and in particular of the sizeable drag increase which it usually produces. 
Accordingly, \citet{roshko1965} show that the length of the recirculation region $L_R$, measured as the streamwise distance between the separation point and the reattachment point, is the characteristic length scale of the \textit{reduced} streamwise pressure distributions past a wide set of different bluff bodies, at least if separation is geometrically fixed.
Since, by definition, $L_R$ is the scale of shear layer development, it is then generally admitted that interacting with the shear layer to artificially tune $L_R$ might be an effective strategy to change the pressure distribution in separating/reattaching flows, and hence to control drag (\citet{chun1996}, \citet{berk2017}, \citet{stella2018}).
Unfortunately, drag control solutions based on this approach have proven hard to scale up from laboratory models to full-size industrial applications. This problem appears to be largely linked to the great sensitivity of the spreading the separated shear layer to a number of flow and geometry parameters, such as free-stream turbulence (\citet{adamsJohnstonPart2}), the shape of the bluff body (\citet{ruck1993}) or, to some extent, its expansion ratio $ER$ (\citet{nadge14}).
In this instance, the boundary layer upstream of separation, when present, deserves particular attention, because it provides the initial conditions of shear layer development (for example, see the discussion of momentum thickness $\theta$ in \citet{chun1996}). As such, it can affect the separating/reattaching flow in many different ways, one classical example being its laminar/turbulent state (\citet{armaly1983}).
Interestingly, the incoming boundary layer appears to have macroscopic effects on the streamwise pressure distribution induced by a separated flow. \citet{tani1961}, \citet{westphal1984} and in particular \citet{adamsJohnstonPart1} show that pressure recovery at reattachment depends on the ratio $\delta_e/h$, where $\delta_e$ is the full thickness of the boundary layer at separation and $h$ is the characteristic cross-stream scale of the bluff body (typically, the height of the BFS). 
More in details, when $\delta_e/h > 0.3$ the streamwise wall-pressure distribution progressively deviates from its pseudo-universal form observed by \citet{roshko1965}: the maximum reduced wall-pressure coefficient $C_{p,max}$ decreases for increasing values of $\delta_e/h$, eventually reaching a minimum value imposed by $ER$.
These results appear to be consistent with some of the key concepts of the theory of \citet{nash1963}, which predicts that the wall-pressure coefficient at reattachment $C_{p,r}$ should decrease as the thickness of the shear layer at separation increases.
In spite of the difference between $C_{p,r}$ and $C_{p,max}$, such agreement suggests that the incoming boundary layer influences the initial development of the separated flow. Depending on the value of $\delta_e/h$, this \textit{footprint} might be more or less persistent, and possibly propagate up to reattachment.
In other words, separating/reattaching flows appear to generally depend on both $h$ and $\delta_e$, with the relative strength of the two characteristic length scales changing across the velocity field (see for example \citet{songEaton2003,songEaton2004}). 
This strongly suggest that, as other multi-scale flows such as plane wakes(\citet{wygnanski1986}), separating/reattaching flows cannot be considered fully self-similar in a general way, unlike free shear layers (\citet{popeTurbulentFlows}).

Lack of self-similarity might have far reaching consequences, because it questions the common assumption that assimilates the separated shear layer to a free shear layer (see \citet{dandois07} and references therein). 
This has proven a useful hypothesis in the inverstigation of separating/reattaching flows. Among other advantages, indeed, it allows us to approximate the cross-stream velocity profile with an error function (\citet{chapman1958,tanner1973}) and hence to provide a scaling for the main mean shear component $\partial U/\partial y$. 
This is a very important result, because mean shear has a key role in amplificating shear layer instabilities, in enhancing turbulent production and in many more (often detrimental) phenomena that are usually companions of separation.
In this respect, lack of self-similarity makes the analysis of these behaviours much harder, because it implies that, for $\delta_e/h > 0.3$, the scaling of $\partial U/\partial y$ also depends on $\delta_e$ and changes in the streamwise direction in a non-trivial way.
For these reasons, investigating the nature of the influence of $\delta_e/h$ at separation seems of great theoretical and practical interest, with possible implications for modeling of separating/reattaching flows and their control in full-scale applications.
In this work, we contribute to this effort by studying the effect of the parameter $\delta_e/h$ on a prototypical separating/reattaching flow.

The first issue to be addressed relates to our capability in identifying and assessing the footprint of the BL. One classical approach might rely on the very lack of self-similarity of the flow. In this view, the profile of, say, mean streamwise velocity is expected to progressively mutate from the log-law typical of boundary layers, to the error function profile that is characteristic of free shear layers. Then, identifying the footprint of the boundary layer comes down to mapping the regions of the flow in which $\delta_e$ dominates the scaling laws.
This kind of analysis can be attempted with some success (\citet{songEaton2004}), but local scaling changes are more visible if the characteristic scales of the flow are sizeably different. Unfortunately, results reported by \citet{adamsJohnstonPart1} suggest that our analysis is most relevant when $\delta_e$ and $h$ are similar (e.g. $\delta_e/h \in \left(0.3,\,1\right)$): then, directly investigating local scaling of velocity profiles is not an efficient tool to track the footprint of the boundary layer.
In this work we propose an original approach to solve this problem, based on the analysis of mass entrainment.
Mass entrainment has the major advantage of being an integral quantity: it does not rely on self-similarity assumption, or any local $\delta_e/h$ effects, and gives a global picture of the footprint of the boundary layer on the separated flow.
In addition, it is well known that mass entrainment drives the growth of turbulent boundary layers (\citet{chauhan14B,chauhan14A}) as well as the development of free shear layers (\citet{popeTurbulentFlows}). In their recent work, \citet{stella2017} quantitatively show that this is also the case in a separated shear layer. 
Anyway, boundary layers and shear layers grow (i.e. entrain external fluid) in sizeably different ways: then, mass entrainment is also likely to be a powerful tracer of differences between the two categories of flows.
A further consideration in favour of our approach stems from the comparison of free shear layers and separated shear layers.
Generally speaking, these flows differ for their geometrical boundary conditions and, depending on $\delta_e/h$, for their initial conditions. However, the role of mass entrainment in their development is similar (\citet{stella2017}). This suggests that mass entrainment might be a robust descriptor of the physical behaviour of a separated flow, regardless to the value of $\delta_e/h$ and hence to the accuracy of the free shear layer approximation.
Significantly, the recent papers by \citet{berk2017} and \citet{stella2018} indicate that this might even be the case if the separated shear layer is forced with an external control action.
In the light of these findings, mass entrainment stands out as a very promising tool to identify the footprint of the incoming boundary layer.

Of course, mass entrainment in turbulent flows is not a new topic. In this respect, many works have highlighted the importance of the Turbulent/Non-Turbulent Interface (TNTI) in transfer of mass, momentum and energy from the free-stream to the turbulent region of the flow. Research has focused on canonical turbulent flows such as jets (\citet{ww09MomTra}, \citet{daSilva2011}), wakes (\citet{bisset2002}) and in particular turbulent boundary layers (\citet{CK1955}, \citet{fiedlerHead1966}, \citet{hedleyKeffer1974A}, \citet{chauhan14B}, \citet{chauhan14B}, \citet{borrell2016} among others).
A first effort to investigate the TNTI in non-canonical flows is reported by \citet{stella2017}, suggesting that some of the lessons learned on canonical flows can be directly extended to the TNTI of separated flows. 
Some aspects of the instantaneous, local behaviour of the TNTI are still under debate, in the first place concerning the nature of its dominant transfer mechanism (see for example \citet{CK1955}, \citet{townsend1966}, \cite{taveira2013} and \citet{mistry2016}). Anyway, it is generally agreed that the \textit{statistical} behaviour of the TNTI respects flow self-similarity.
In particular, it is known since the seminal work of \citet{CK1955} that in a turbulent boundary layer the instantaneous TNTI location above the wall approximately follows a gaussian distribution, scaled by the thickness of the boundary layer $\delta$ (\citet{chauhan14A} and references therein).
These findings can be very useful for our present investigation. In fact, although the incoming boundary layer might be subjected to a weak pressure gradient (see \cite{kourta15} among others), it does not seem unreasonable to assume that its TNTI will still be approximately gaussian distributed, and scaled by $\delta$. 
It can also be expected that, as the separated flow departs from self-similarity, this characteristic gaussian TNTI \textit{signature} will progressively fade into a different distribution.
Then, the first objective of the present study is to investigate the local distribution of the TNTI over a separated flow. This should allow to extend the work of \citet{stella2017} on the TNTI of a separating/reattaching flow, while contributing at identifying the footprint of the incoming boundary layer. 

Our second objective consists in investigating how such footprint modifies the development of the flow in the region downstream of separation. In particular, we are interested in understanding how the behaviour of the shear layer is affected by the variation of $\delta_e/h$. This certainly is a vast subject, that cannot be exhausted in a single study. Anyway, a question of primary importance that can be addressed with reasonable effort concerns the scaling of the main mean shear component $\partial U/\partial y$. Indeed, as already stated, the introduction of $\delta_e$ as a second characteristic scale of the separated flow might have far reaching consequences on how the shear layer shapes many properties of the entire flow. 

As a third contribution, we use mass entrainment to investigate whether $\delta_e/h$ affects the free shear layer analogy.
Under this hypothesis, indeed, the mean spreading rate of the separated shear layer is considered proportional to the sum of mean mass entrainment rates at its boundaries \citet{popeTurbulentFlows}. This has been verified by \citet{stella2017} even with a relatively high value of $\delta_e/h$, but the possible effects of the incoming boundary layer on entrainment rates has not yet been analysed. Anyway, if $\delta_e/h$ modifies the velocity field (e.g. $\partial U/\partial y$), it can be expected that it might also have a sizeable impact on the scaling of mass entrainment rates, and possibly on the free shear layer analogy. 

The paper is structured as follows: § \ref{sec:expSetUp} presents the experimental set-up; § \ref{sec:competition} deals with TNTI statistics and with the identification of the footprint of the boundary layer; the development of the separated shear layer and its scaling are treated at § \ref{sec:sepShLayAnalysis}; § \ref{sec:massEntrShLayer} analyses the relationship between shear layer growth and mass entrainment; conclusions are given at § \ref{sec:conclu}.
In the remainder of the paper, we will adopt the Reynolds decomposition of the velocity field and its standard notation. For example, the instantaneous streamwise velocity component $u$ will be expressed as:
\begin{equation}
u = U + u^\prime,
\end{equation}
where $U$ and $u^\prime$ are the mean and fluctuating streamwise velocities, respectively. The same convention applies to the wall-normal velocity component $v$. The symbol $^*$ is used to indicate normalisation of lengths on $L_R$, or of mass fluxes on $\rho U_\infty h$, where $\rho$ is density of air and $U_\infty$ is free-stream velocity.

\section{Experimental set-up}\label{sec:expSetUp}
\begin{figure}
\setlength{\unitlength}{1cm}
\begin{center} 
\includegraphics[width=0.9\textwidth]{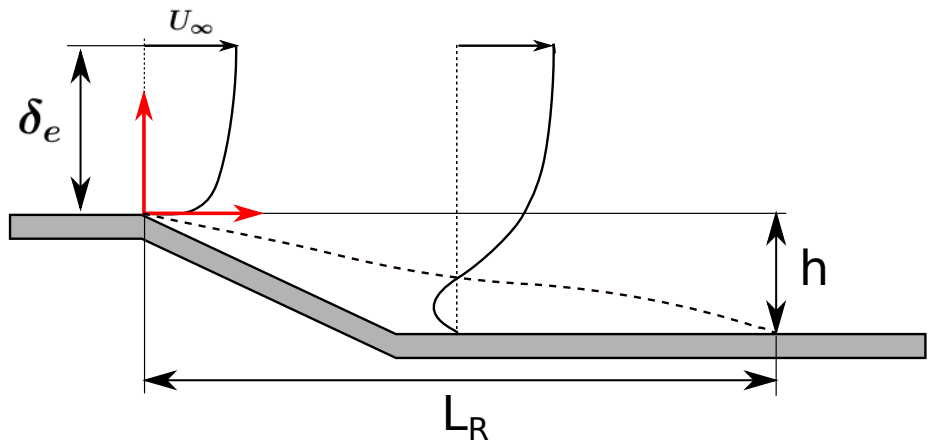}
\end{center}
\caption[Outline of the flow]{Outline of the mean separating/reattaching ramp flow investigated in this study. $X$ and $Y$ are respectively the streamwise axis and the wall-normal axis of a reference system centered on the salient edge at which the flow separates. \protect\dashedrule Recirculation Region Interface. The three characteristic length scales of the flow ($\delta_e$, $h$ and $L_R$) are also reported. $U_\infty$ is measured above the upper edge of the ramp.}
\label{fig:rampeGeneric}
\end{figure}

The reference geometry for this research is a descending, \SI{25}{\degree} ramp that causes the massive separation of an incoming turbulent boundary layer (see figure \ref{fig:rampeGeneric}).
This section presents the two experimental models as well as the measurements techniques used in this study.
\subsection{Experimental models}\label{sec:expeModel} 
Experiments were carried out on two geometrically similar ramps, spanning two different step heights but with essentially similar values of $\delta_e$. This allows to study the effect on the flow of sizeably different values of the ratio $\delta_e/h$. 
The first experimental model is the so-called \textit{GDR ramp}, which was used in previous studies such as \citet{debien14} and \citet{kourta15}. The reader is referred to these works for a complete description of the model, the main properties of which are summarised in table \ref{tab:ramps_comp}.
The GDR step height is $h = $ \SI{100}{\milli\metre}.
The second experimental model was already presented in details in \citet{stella2017}. For simplicity, in the remainder of this paper it will be indicated as the \textit{R2 ramp}. The R2 ramp has a step height $h =$ \SI{30}{\milli\metre}, but values of the expansion ratio $ER$ and of the aspect ratio $AR$ are comparable to those of the GDR ramp (see table \ref{tab:ramps_comp}). On the contrary, the ratio $\delta_e/h$ is about three times higher than on the GDR ramp.
Together, the two experimental models allow us to cover almost \num{1.5} decades of the similarity parameter $\Rey_h = U_\infty h/ \nu$, where $U_\infty$ is a reference velocity and $\nu$ is the kinematic viscosity of air.

\begin{table}
\centering
\begin{tabular}{@{}cccccccc@{}}
 && $h/$\si{\milli\metre} & $AR$ & $ER$ & $\Rey_h/$\num{e4} & $\delta_e/h$ & $\sqrt{u^{\prime 2}}/U_\infty$ [\si{\percent}]\\
\midrule
GDR && 100 & 20 & 1.11 & \numrange{7}{35} & $\approx 0.3$ & <0.3 \\
R2 && 30 & 17 & 1.06 & \numrange{1}{8} & $\approx 0.85$ & <0.25 \\
\end{tabular}
\caption{Geometry-based parameters of the GDR ramp, and of the R2 ramp used in \citet{stella2017}. $h$ is the height of the step, and $AR$ and $ER$ are the aspect ratio and the expansion ratio, respectively. The ratio $\sqrt{u^{\prime 2}}/U_\infty$, evaluated within the free stream, is used to measure residual turbulent intensity in the two facilities.}
\label{tab:ramps_comp}
\end{table} 

\subsection{Measurement devices}
The investigation of the massive turbulent separation is mainly based on Particle Image Velocimetry (PIV). Since the mean flow is bidimensional (see \citet{kourta15} and \citet{stella2017}), 2D-2C PIV is relevant for the purposes of this study.
On the GDR ramp, PIV images are obtained with two LaVision VC-Imager cameras, synchronised with a double pulse Nd:Yag laser (wavelength \SI{532}{\nano\metre}, rated $2 \times 200$ \si{\milli\joule}). Each camera is equipped with a Nikon Nikkor 105 lens, yielding an image resolution of about \SI{118}{\micro\metre\per\pixel} on a field of view of $4.6h \times 2h$. 
The flow is seeded with olive oil droplets of mean diameter $d_p = $ \SI{1}{\micro\metre}. Their characteristic time response is low enough for them to accurately trace all scales of the flow (see \citet{stella2017}).
For each $\Rey_h$, \num{2000} particle image pairs are recorded at midspan, with an aquisition rate of \SI{2}{\hertz}. Then, image pairs are correlated with the multipass, FFT algorithm of the Davis 8.3 software suite. The size of the final correlation window is \SI[parse-numbers = false]{32 \times 32}{\square\pixel} with \SI{50}{\percent} overlapping, yielding a space resolution $\Delta/h \approx $ \num{0.03}, which is adequate for an investigation of the mean field. After correlation, the vector fields yielded by the two cameras are merged, for a total field of view of $6h \times 2h$.
On the R2 ramp, PIV images are recorded with one LaVision VC-Imager camera, equipped with Zeiss \SI{50}{\milli\metre} ZF Makro Planar T* lens, which provides a camera resolution of  \SI{78}{\micro\metre\per\pixel} and an exploitable field of view of 6h x 2.5h. Laser setting were identical as on the GDR ramp.
For each tested $\Rey_h$, the R2 ramp database provides \num{2} fields of view, partially overlapping. For each field of view, a set of \num{2000} PIV images is available. Instantaneous images of differents sets are not correlated, but field statistics can be merged to give a total field of view of approximately $9h \times 2.5h$, covering the entire mean recirculation region. PIV images are correlated with the multipass, GPU direct correlation algorithm of the LaVision Davis \num{8.3} software suite. The size of the final interrogation window is \SI[parse-numbers = false]{16 \times 16}{\square\pixel}, with \SI{50}{\percent} overlapping. The spatial resolution is $\Delta/h \approx $ \num{0.04}.

\begin{figure}
\setlength{\unitlength}{1cm}
\begin{center} 
\includegraphics[width=0.9\textwidth]{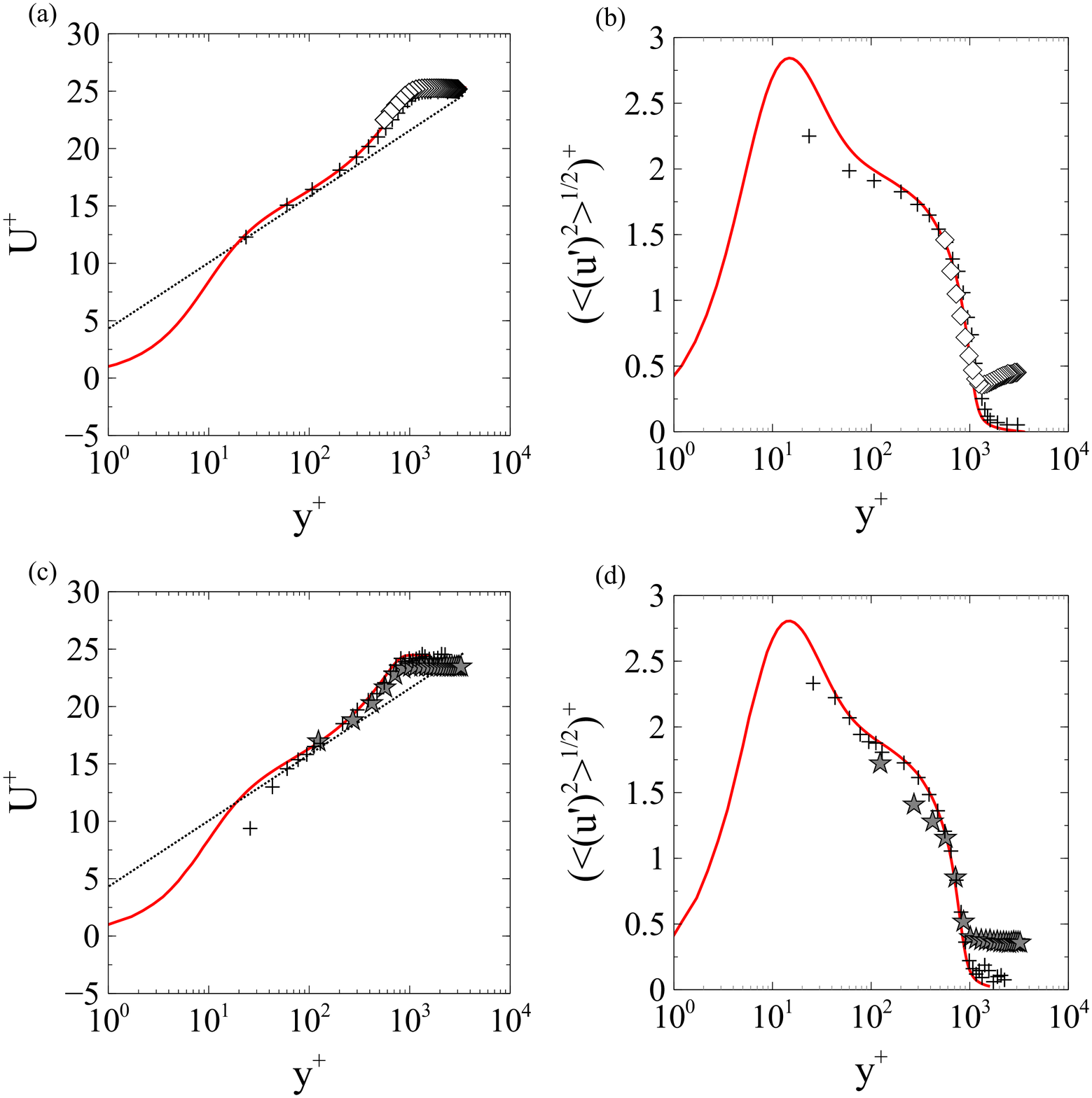}
\end{center}
\caption[BL profiles]{Reference boundary layer profiles at $x/h = -9$ in internal units. (a) $U^+$ and (b) $\sqrt{\langle (u^\prime)^2\rangle}^{\,+}$ for the R2 ramp at $\Rey_\theta =$ \num{3262}; (c) $U^+$ and (d) $\sqrt{\langle (u^\prime)^2\rangle}^{\,+}$ for the GDR ramp at $\Rey_\theta =$ \num{3617}. Symbols: + hotwire measurements; \graylozenge PIV data from the R2 ramp auxiliary field; \graystar PIV data from the GDR ramp auxiliary field. {\color{red}\protect\rule[0.5ex]{1cm}{1pt}} DNS at $\Rey_\theta =$ \num{2537} as given in \citet{DNS10}. For clarity, only one point every three is reported for each dataset.}
\label{fig:BLayer_plus}
\end{figure}

The thickness of the boundary layer at separation $\delta_e$ is measured with a single-component hot-wire probe (Dantec 55P15), driven in constant-temperature mode by a Dantec Streamline 90N10 frame. The sensing length of the probe is $\ell_w = $ \SI{1.25}{\milli\metre}. A discussion of the filtering effect due to $\ell_w$ is provided by \citet{philip13}.  
It is stressed that the value of $\delta_e$ considered here is the \textit{full thickness} of the boundary layer, i.e. the distance at which $U \approx U_\infty$. 
The turbulent state of the incoming boundary layer is often evaluated with the parameter $\Rey_\theta = \theta U_\infty/\nu$, where $\theta$ is momentum thickness (see for example \citet{songEaton2004}). To allow comparison with \citet{stella2017}, $\Rey_\theta$ is assessed from hot-wire measurements at a reference section placed at $x/h = -9$. In addition, auxiliary sets of \num{2000} PIV images of the incoming boundary layer are recorded at this reference section. For both ramps, characteristics of the PIV set-up and correlation settings are the same as detailed above for the main PIV fields. This gives $\Delta/\delta \approx 0.09$ on the GDR ramp and $\Delta/\delta \approx 0.04$ on the R2 ramp, where $\delta$ is the thickness of the boundary layer at $x/h = -9$. Figure \ref{fig:BLayer_plus} shows that velocity profiles of the reference turbulent boundary layer, obtained both from hot-wire and PIV measurements, agree with DNS data by \citet{DNS10} sufficiently well to confidently consider larges-scale properties such as $\theta$.

\subsection{Estimating the recirculation length}\label{sec:massiveSep}
\begin{figure}
\setlength{\unitlength}{1cm}
\begin{center} 
\includegraphics[width=\textwidth]{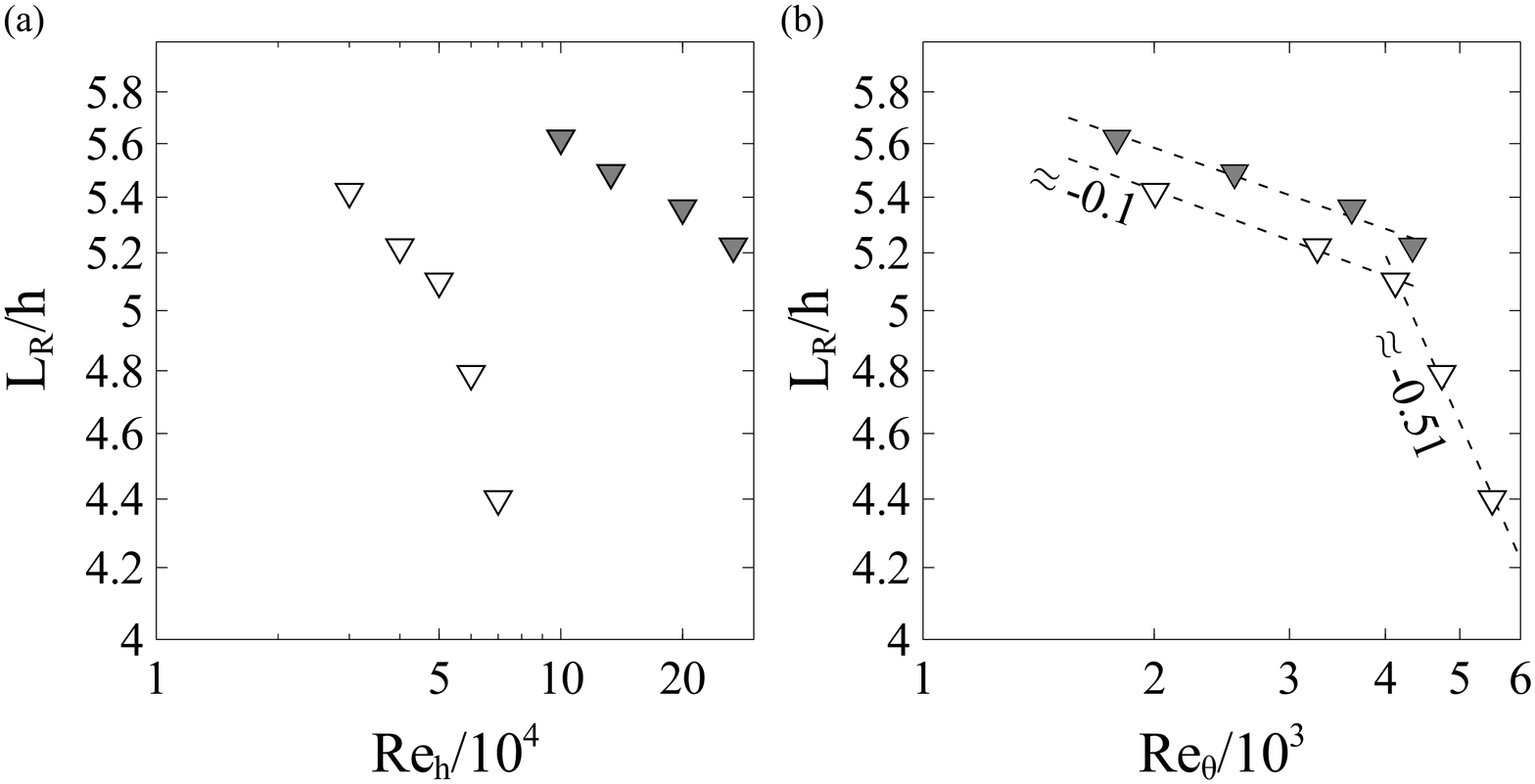} 
\end{center}
\caption{Evolution of $L_R/h$ with (a) $\Rey_h$ and (b) $\Rey_\theta$. Symbols: $\triangledown$ R2 ramp; $\color[rgb]{0.5,0.5,0.5}{\blacktriangledown}$ GDR ramp; \protect\dashedrule power-law fits (in the sense of least-mean square).}
\label{fig:LR_RE_THETA}
\end{figure}

Since the mean topology of the separated flow is substantially comparable across experiments, figure \ref{fig:rampeGeneric} reports a generic sketch of the mean streamwise velocity field. 
The incoming boundary layer separates at the upper edge of the ramp, giving origin to the separated shear layer and the recirculation region.
The external boundary of the recirculation region is the mean separation line. For consistency with \citet{stella2017}, we will indicate it as the Recirculation Region Interface (RRI). The RRI is defined by the isoline $U=0$ on the mean streamwise velocity field (see \citet{kourta15}). In principle, the reattachment point can be identified as the last point of the mean RRI (see for example \citet{le1997} or \citet{kourta15}). Unfortunately, in most of our PIV datasets a thin region in close proximity of the wall was unexploitable, due to laser reflections. Then, $L_R$ was estimated as follows: the RRI was approximated with two polynomials, joint at $x/h \approx 2.5$ (see figure \ref{fig:rampeGeneric}) and conditioned as to have a continuous first derivative. Then, this polynomial RRI was extrapolated to $y/h = -1$. 
Values of $L_R/h$ so obtained are listed in table \ref{tab:flow_props}. 
Estimated $L_R/h$ of both ramps appear to scale with $\Rey_\theta$, according to a relationship that \citet{stella2017} modelled with a power-law, in the form:
\begin{equation}
	L_R/h \sim \Rey_\theta^m,
	\label{eq:REm}
\end{equation}
where $m \approx -0.1$ for $\Rey_\theta < \Rey_{\theta,c}$ and $m \approx -0.55$ for $\Rey_\theta > \Rey_{\theta,c}$. Even if the cause of the change of exponent was not identified in that study, the critical value $\Rey_{\theta,c}$ was evaluated to approximately \num{4100}. 
As shown in figure \ref{fig:LR_RE_THETA}, $\Rey_\theta$ appears to be more adapted at scaling $L_R/h$ across experiments than $\Rey_h$, at least for $\Rey_\theta < \Rey_{\theta,c}$. For this reason, in the remainder of the paper we will systematically consider $\Rey_\theta$ as the relevant Reynolds number of the flow. 

\begin{table}
\centering
\begin{tabular}{@{}cccccccccccc@{}}
$\Rey_h/$\num{e4} && \num{3} & \num{4} & \num{5} & \num{6} & \num{7} && \num{10} & \num{13.3} & \num{20} & \num{26.7} \\     
\midrule
$\delta_e/h$ && 0.92 & 0.86 & 0.82 & 0.82 & 0.91 && 0.34 & 0.28 & 0.33 & 0.29 \\
 && & & & & && & & & \\
$u_\tau/[ms^{-1}]$ && 0.68\ding{61} & 0.78 & 0.93 & 1.10 & 1.25 && 0.7\ding{61} & 0.86\ding{61} & 1.12 & 1.57 \\
$\Rey_\tau$ && 1270 & 1310 & 1750 & 2130 & 2646 && 840 & 1090 & 1456 & 1622 \\
$\Rey_\theta$ && 2006 & 3262 & 4122 & 4738 & 5512 && 1788 & 2547 & 3617 & 4340 \\ 
$H_{12}$ && 1.43 & 1.43 & 1.40 & 1.40 & 1.37 && 1.57 & 1.57 & 1.42 & 1.40 \\
 && & & & & && & & & \\
$L_R/h$ && 5.42 & 5.22 & 5.1 & 4.79 & 4.4 && 5.62 & 5.49 & 5.37 & 5.22\\
\end{tabular}
\caption[Characteristic parameters of the flow over the two ramps.]{Characteristic parameters of the flow over the two ramps. The ratio $\delta_e/h$ is computed at $x/h \approx 0$, as the distance from the wall where $U \approx U_\infty$. 
Other boundary layer properties are measured at a reference section x/h = \num{-9}, at which the pressure gradient is zero (see \citet{stella2017}). Friction velocity $u_\tau$ is obtained with the composite profile of \citet{chauhan09}, with the exception of those marked with the symbol \ding{61}, that were computed with the Clauser chart method (see \citet{wei_Clauser2005}, among others). It is $\Rey_h = U_\infty h/\nu$, $\Rey_\tau = \delta u_\tau/\nu$ and $\Rey_{\theta} = U_\infty \theta/\nu$. $H_{12}$ is the shape factor ($\equiv \delta_1/\theta$, where $\delta_1$ is the displacement thickness). 
$L_R/h$ is computed from a polynomial fit of the mean RRI, detected on the mean streamwise PIV velocity field (see § \ref{sec:massiveSep}).}
\label{tab:flow_props}
\end{table} 

\section{A footprint of the incoming boundary layer}\label{sec:competition}
The $\delta_e/h$ effect observed at reattachment (\citet{adamsJohnstonPart1}) suggests that the incoming boundary layer influences the flow after separation. 
If this is so, it can be expected that the separated flow contains a distinctive footprint of the incoming boundary layer, that possibly survives up to reattachment. It seems then convenient to start our investigation by identifying such footprint, and by showing that its strenght depends on $\delta_e/h$.
In order to do so, we use the statistical distribution of the Turbulent/Non-Turbulent Interface (TNTI) as a tracer, to highlight changes of flow properties in the streamwise direction. This choice is based on several TNTI characteristics that appear to be well suited to our purposes. 
Firstly, the TNTI is an inexpensive tracer. Indeed, the TNTI can be detected even on simple 2D2C PIV fields, with no need of extra experimental or numerical efforts (\citet{chauhan14A}). 
Secondly, in our flow the TNTI exists on the entire velocity field, which allows simple assessment of the streamwise evolution of the flow (\citet{stella2017}). Finally and most importantly, the statistical distribution of the TNTI provides a reliable boundary layer \textit{signature}, to which we can compare the separated flow under study. In particular, we remind that in a turbulent boundary layer the instantaneous TNTI location above the wall approximately follows a gaussian distribution, scaled by $\delta$ (see references at § \ref{sec:intro}). 

In the following subsections, we investigate the TNTI distribution in the separated flow under study, to determine wheather the gaussian form typical of the incoming boundary layer survives to separation. It is expected that TNTI properties will remain similar to those of the incoming boundary layer, as long as the latter has a dominant influence on the separated flow. Comparison of the two ramps allows to highlight the effects of the parameter $\delta_e/h$. The TNTI is detected following the method proposed by \citet{chauhan14A} and adapted to separated flow by \citet{stella2017}. Its main steps are reminded hereafter.

\subsection{Detection of the TNTI}
Following \citet{daSilva2014} and \citet{chauhan14B,chauhan14A}, the instantaneous TNTI can be identified with a threshold on the field of a dimensionless turbulent kinetic energy, defined as:
\begin{equation}
\tilde{k} = \frac{100}{9(U_{\infty}^2 + V_{\infty}^2)}\sum_{m,n=-1}^{1}[(u_{m,n} - U_{\infty})^2+(v_{m,n} - V_{\infty})^2].
\label{eq:kinEn} 
\end{equation}
In eq. \ref{eq:kinEn}, $\tilde{k}$ is locally averaged on a kernel of side equal to \num{3} vectors, to smooth out PIV noise. The indexes $m$ and $n$ allow to iterate on the two dimensions of the kernel.
The threshold value $\tilde{k}_{th}$ is computed iteratively from instantaneous PIV snapshots of the incoming boundary layer, captured on the auxiliary PIV fields. Based on several analysis of TNTI statistics in turbulent boundary layers (see \citet{CK1955} and \citet{chauhan14intrcy} among others), $\tilde{k}_{th}$ is chosen as the smallest $\tilde{k}$ value for which the following condition is verified:
\begin{equation}
Y_T + 3\sigma_T \approx \delta,
\label{eq:TNTIdet_kConvCrit} 
\end{equation}
where $Y_T$ is the mean wall-normal TNTI position, $\sigma_T$ is its standard deviation and $\delta$ is estimated with the composite boundary layer profile conceived by \citet{chauhan09}.
$\tilde{k}_{th}$ strongly depends on free-flow turbulence and PIV noise, but it is relatively insensitive to $\Rey_\theta$. In the case of the R2 ramp, one can consider $\tilde{k}_{th} = 0.365$. As for what concerns the GDR ramp, a reference threshold value $\tilde{k}_{th} \approx 0.21$ can be retained for all $\Rey_\theta$. The mean TNTI is identified simply, by detecting the $\tilde{k}_{th}$ isoline on the mean $\tilde{k}$ field. Values of $Y_T/\delta$ and $\sigma_T/\delta$ corresponding to $\tilde{k}_{th}$ are listed in table \ref{tab:TNTIdet_stat_TNTIPos_BL_PIV}. Higher order statistics, defined in the following subsection, are also listed for future reference.
\begin{table}
\centering
\begin{tabular}{@{}lccccccc@{}}
\multicolumn{2}{l}{} & $\Rey_\theta$& $\Rey_\tau$ & $Y_T/\delta$ & $\sigma_T/\delta$ & $Sk_T$ & $Kt_T$\\
\midrule
\citet{chauhan14A} 								&						& - 	& 14500 & 0.67 & 0.11 & - 		& - \\
\multirow{5}{*}{Present study R2}  				& \ldelim\{{5}{15pt} 	& 2006 	& 1300 	& 0.64 & 0.12 & 0.36 	& 3.32 \\
												&						& 3262 	& 1310 	& 0.60 & 0.13 & 0.34 	& 3.12 \\
												&						& 4122 	& 1750 	& 0.61 & 0.13 & 0.36 	& 3.12 \\
												& 						& 4738 	& 2130 	& 0.61 & 0.13 & 0.38 	& 3.17 \\
												& 						& 5512 	& 2646 	& 0.62 & 0.12 & 0.34 	& 3.16 \\
\multirow{4}{*}{Present study GDR}  			& \ldelim\{{4}{15pt} 	& 1788 	& 874 	& 0.62 & 0.13 & 0.32 	& 3.58\\
												&						& 2545	& 1128 	& 0.62 & 0.13 & 0.34 	& 3.20\\
												&						& 3617 	& 1490 	& 0.67 & 0.11 & 0.23 	& 3.10\\
												& 						& 4340 	& 1883 	& 0.63 & 0.12 & 0.25 	& 3.00\\
\end{tabular}
\caption[PIV-based statistics of TNTI position in the boundary layer]{Comparison of statistics of TNTI position in the boundary layer. $Y_T$ is the mean TNTI position above the wall and $\sigma_T$ is its standard deviation. $Sk_T = \mu_{T3}/{\mu_{T2}}^{3/2}$ and $Kt_T = \mu_{T4}/{\mu_{T2}}^{2}$ are the skewness and the kurtosis coefficients of the TNTI distribution, respectively (see § \ref{sec:statDefs}). Expected values for a normal distribution are $Sk_T = 0$ and $Kt_T = 3$.}
\label{tab:TNTIdet_stat_TNTIPos_BL_PIV}
\end{table}
\subsection{Statistical distribution of the TNTI: definitions}\label{sec:statDefs}
\begin{figure}
\setlength{\unitlength}{1cm}
\begin{center} 
\fbox{\includegraphics[width=0.96\textwidth]{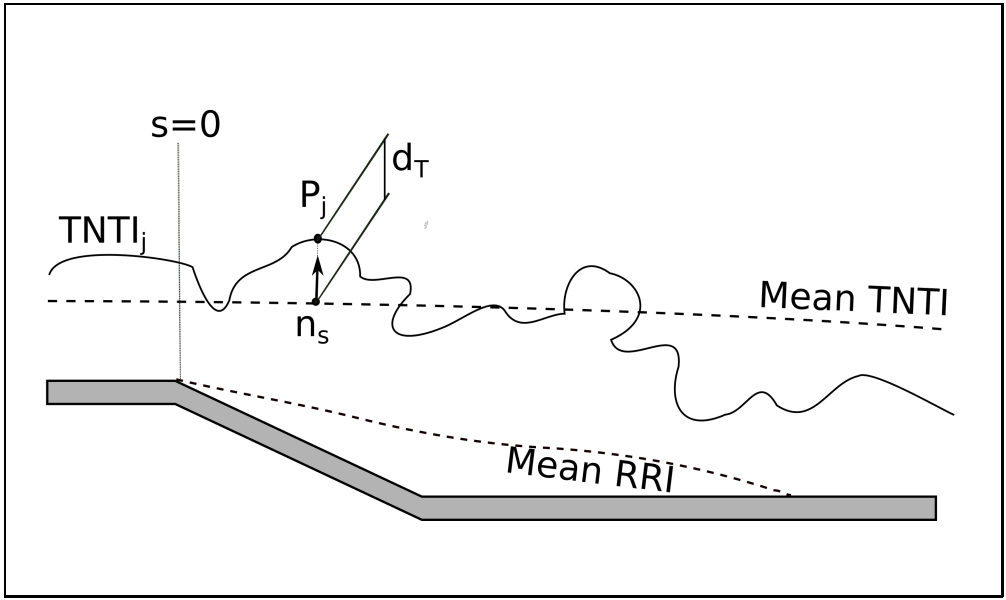}}
\end{center}
\caption{Notations and reference frames for the study of the TNTI distribution in space. $d_{T}(s^*)$ is, the position of the instantaneous TNTI at the curvilinear abscissa $s^*$ (defined on the mean TNTI).}
\label{fig:interf_references}
\end{figure}

For what follows, it is practical to compute the positions of the TNTI in the local frame presented in figure \ref{fig:interf_references}, which generalises the streamwise-wall normal one used for boundary layers (\citet{CK1955}, \citet{chauhan14A}).  
Let us start by considering the mean TNTI. This is a smooth line, which at separation is placed at $Y_{T,e}/\delta_e \approx 0.5$, where $Y_{T,e}$ is the mean TNTI position above the upper edge of the ramp. After separation, the mean TNTI gently curves toward the wall. We define a curvilinear abscissa \textit{s} along the mean TNTI. For simplicity, the origin of $s$ is placed at $x=0$. Let also $\boldsymbol{n}_{s}$ be a normal to the mean TNTI. The instantaneous TNTI location for any given $s$ is defined by $d_{Tj}(s)$, the signed distance (positive up) along $\boldsymbol{n}_{s}$ between the j-th instantaneous TNTI and the mean TNTI. The probability density function of $d_{Tj}(s)$ is then $N_{T}(s)$.
To characterise $N_{T}(s)$, we will consider its standard deviation $\sigma_T$, its skewness coefficient $Sk_T = \mu_{T3}/{\mu_{T2}}^{3/2}$ and its kurtosis coefficient $Kt_T = \mu_{T4}/{\mu_{T2}}^{2}$. In these expressions, $\mu_{Tn} = E\left(d_T - E\left(d_T\right)\right)^n$ is the $n^{th}$-order central moment of $N_T$, and $E$ is the expected value. According to \citet{cintosun2007}, $\sigma_T$ should be related to the characteristic large scale of the flow. Then, it should be $\sigma \sim \delta$ in those regions in which the footprint of the incoming boundary layer is strong. The coefficients $Sk_T$ and $Kt_T$ carry information on the shape of $N_{T}(s)$, respectively on its symmetry and on the size of its tails (i.e. on the presence of outliers, see \citet{westfall2014}). For a gaussian distribution, it is $Sk_T=0$ and $Kt_T =3$.
\subsection{Streamwise evolution of the statistical distribution of the TNTI}
\begin{figure}
\setlength{\unitlength}{1cm}
\begin{center} 
\includegraphics[width=0.9\textwidth]{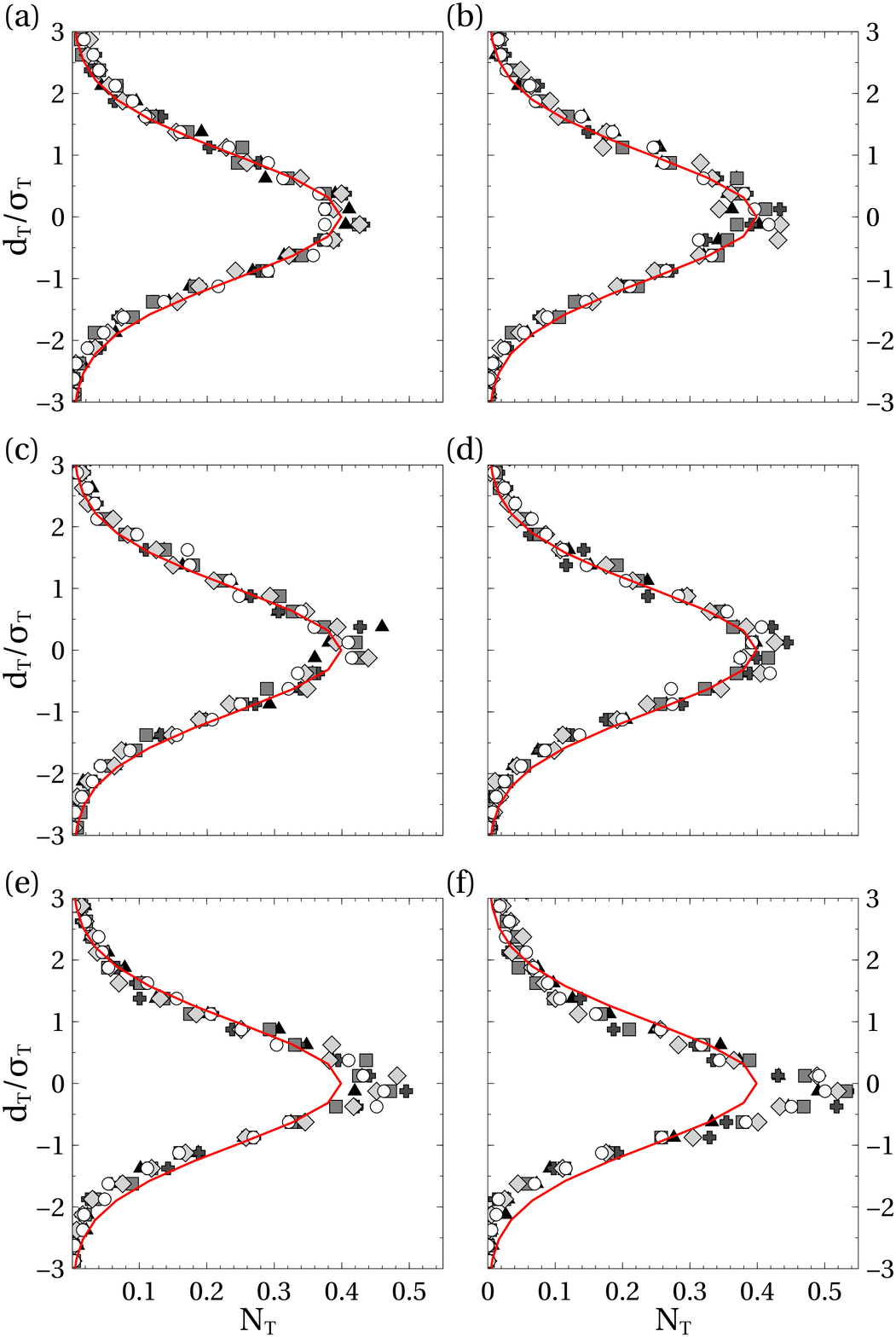}
\end{center}
\caption[Normalised TNTI distribution on the R2 ramp]{Normalised TNTI distribution on the R2 ramp. (a) $x^* = 0.2$; (b) $x^* = 0.4$; (c) $x^* = 0.6$; (d) $x^* = 0.8$; (e) $x^* = 1.1$; (f) $x^* = 1.4$. Symbols: $\bigcirc$ $\Rey_\theta=$ \num{2006}; {\contour{black}{$\color[rgb]{0.75,0.75,0.75}{\blacklozenge} \,$}} $\Rey_\theta=$ \num{3262}; {\contour{black}{$\color[rgb]{0.5,0.5,0.5}{\blacksquare} \,$}} $\Rey_\theta=$ \num{4122}; {\color[rgb]{0.25,0.25,0.25}\ding{58} } $\Rey_\theta=$ \num{4738}; $\blacktriangle$ $\Rey_\theta=$ \num{5512}. {\color[rgb]{1,0,0}{\protect\solidrule}} gaussian distribution.}
\label{fig:R2_TNTI_PDF}
\end{figure}

\begin{figure}
\setlength{\unitlength}{1cm}
\begin{center} 
\includegraphics[width=0.9\textwidth]{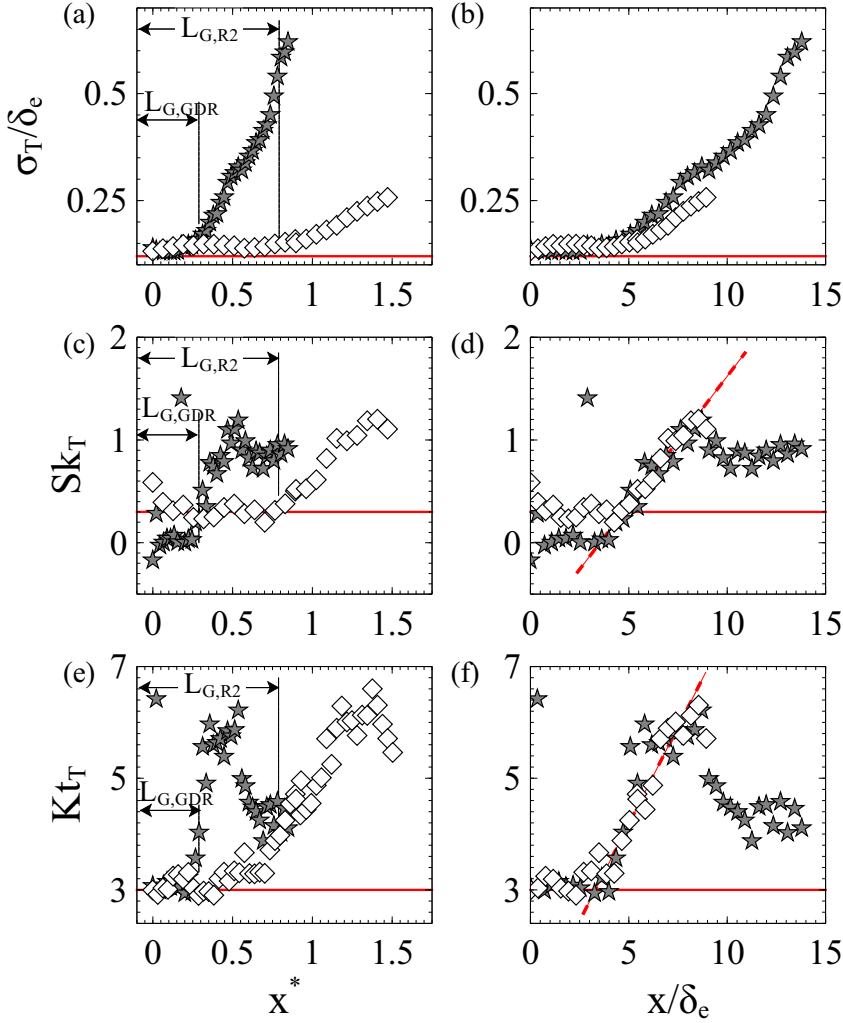}
\end{center}
\caption[TNTI stats]{Streamwise evolution of TNTI statistics, represented with different normalisation: (a) and (b) $\sigma_T/\delta_e$; (c) and (d) skewness coefficient $Sk_T = \mu_{T3}/{\mu_{T2}}^{3/2}$; (e) and (f) kurtosis coefficient $Kt_T = \mu_{T4}/{\mu_{T2}}^{2}$. Subfigures in the same column share the same streamwise normalisation: (a),(c) and (e) $x/L_R$; (b), (d) and (f) $x/\delta_e$. Symbols: {\protect\graylozenge} R2 ramp ($\Rey_\theta =$ \num{3262}); \graystar GDR ramp ($\Rey_\theta =$ \num{3617}). Solid lines ({\color[rgb]{1,0,0}{\protect\solidrule}}) show reference values for the gaussian TNTI distribution observed in the incoming boundary layer, at the reference section $x/h \approx -9$ (see \ref{tab:TNTIdet_stat_TNTIPos_BL_PIV}): (a) and (b) $\sigma_T/\delta_e \approx 0.12$; (c) and (d) $Sk_T = 0.3$; (e) and (f) $Kt_T = 3$. A dashed line ({\color[rgb]{1,0,0}{\protect\dashedrule}}) is used to highlight common trends.}
\label{fig:TNTI_STATS}
\end{figure}

The form of $N_{T}(s)$ over the R2 ramp is reported in figure \ref{fig:R2_TNTI_PDF}, at several streamwise locations. In a large neighbourhood of the mean separation point, the TNTI seems roughly distributed as a gaussian random variable: the statistical properties of the TNTI that are typical of boundary layers seem then to survive to separation, persisting over a part of the recirculation region. Let us call \textit{gaussian length} (noted $L_G$) this first part of the separated flow. Further downstream, however, $N_{T}(s)$ deviates progressively from a gaussian distribution. $N_{T}(s)$ is more and more skewed: its \textit{inner} (i.e. towards the wall) tail is shortened and the TNTI sample is more concentrated slightly under the mean TNTI. The streamwise evolution of some TNTI properties is not surprising: indeed, unlike canonical flows studied in previous works, the massive separation under investigation is not equilibrated (since the boundary conditions evolve) nor, in general, self-similar (see for example \citet{songEaton2004}). The evolution of $N_{T}(s)$ on the GDR ramp (not reported) is qualitatively comparable to the one observed on the R2 ramp, but $L_G$ is found to be much longer in the latter experiment. The analysis of the streamwise evolution of statistical moments allows to sketch a possible explanation for different $L_G$ magnitudes.
The streamwise evolutions of $N_T$ statistics over both ramps are presented in figure \ref{fig:TNTI_STATS}. For each ramp, all curves collapse together nicely. Then, only R2 data at $\Rey_\theta =$ \num{3262} and GDR data at $\Rey_\theta =$ \num{3617} are compared, because their incoming boundary layers have very similar turbulent states. Figures \ref{fig:TNTI_STATS} (a), (c) and (e), respectively show the evolutions of $\sigma_T/\delta_e$, $Sk_T$ and $Kt_T$ in function of the non-dimensional streamwise coordinate $x^*$.  
It appears that, in a large region downstream of separation, all considered TNTI statistics keep values that are very similar to those measured in the incoming boundary layer, at $x/h \approx - 9$, with substantial agreement among the two ramps.
These quantitative observations confirm that the $\delta$-scaled, gaussian form of $N_T$ persists downstream of separation, over a domain $x \in (0,L_G)$.
According to figure \ref{fig:TNTI_STATS} (a), $L_G$ does not scale with $L_R$. Indeed, it is $L_G^* \approx$ \num{0.8} on the R2 ramp and $L_G^* \approx $ \numrange{0.2}{0.3} in the case of the GDR ramp. Anyway, since it is $h_{GDR}/h_{R2} = 3$ and $L_R/h \approx 5$ for both ramps (at least if $\Rey_\theta$ dependencies are neglected), the \textit{dimensional} value of $L_G$ is approximately constant across experiments.
Interestingly, on both ramps it is $L_G^* \approx \delta_e/h$: since $L_R \sim h$, it is then tempting to put $L_G \sim \delta_e$.
Figure \ref{fig:TNTI_STATS} (b), (d) and (f) appear to support this idea. By normalising the streamwise coordinate on $\delta_e$, indeed, trends of all $N_T$ parameters collapse at least on the entire extent of $L_G^*$. 

As for what concerns the domain $x^* > L_G^*$, the footprint of the incoming boundary layer appears to progressively wane. $N_T$ deviates from a gaussian distribution, as shown by the increasing values of both $Sk_T$ and $Kt_T$ (figure \ref{fig:TNTI_STATS} (c) and (e)). Positive skewness coefficients are compatible with a longer \textit{outer} (i.e. toward the free stream) tail and higher values of $Kt_T$ indicate a stronger presence of outliers. In addition, $\sigma_T$ also increases on both ramps, with approximately linear trends. 
The behaviour of $\sigma_T$ reminds the linear increase of the cross-stream scale of the shear layer (see § \ref{sec:sepShLayAnalysis}), so that it does not seem unreasonable to associate the domain $x^* > L_G^*$ to a certain predominance of the separated shear layer. Anyway, mind that in free shear layers $N_T$ has also been observed to be gaussian (\citet{attili2014}): then, the non-gaussian form found on $x^* > L_G^*$ might be indicative of a transition region, possibly toward a new boundary layer, dominated by the development of the shear layer.
All in all, the analysis of the statistical behaviour of the TNTI highlights a sizeable footprint of the incoming boundary layer on the separated flow. Such footprint dominates the flow on a length $L_G$ downstream of separation, but progressively weakens as the separated shear layer develops. 
More importantly, the parameter $\delta_e/h$ appears to determine the relative strength of the incoming boundary layer with respect to the development of the separated shear layer. The higher is $\delta_e/h$, the more the footprint of the incoming boundary layer is persistent. 

\section{Effects on the development of the separated shear layer}\label{sec:sepShLayAnalysis}
\begin{figure}
\setlength{\unitlength}{1cm}
\begin{center} 
\includegraphics[width=\textwidth]{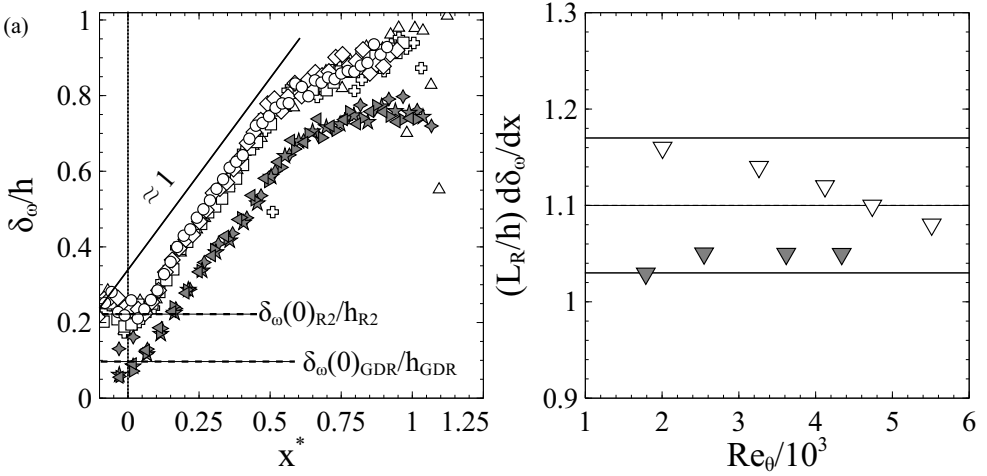} 
\end{center}
\caption[Shear layer growth]{Shear layer scaling. (a) Shear layer development, expressed as the streamwise evolution of vorticity thickness $\delta_\omega$. For sake of clarity, only one in each six points is represented in the case of the R2 ramp, and only one in fifteen in the case of the GDR ramp. The solid line (\protect\solidrule) indicates the unity slope. For reference, the upper edge of the ramp ($x/L_R = 0$) is visualised by a dotted line (\protect\dottedrule) and the corresponding values of $\delta_\omega/h$ by dashed lines (\protect\dashedrule). Symbols for the GDR ramp (grey online): \graytriangleleft $\Rey_\theta = $ \num{1788}; \graytriangleright $\Rey_\theta = $ \num{2547}; \graystar $\Rey_\theta = $\num{3617}; \grayfourstar $\Rey_\theta = $ \num{4340}. Symbols for the R2 ramp (white online) as in figure \ref{fig:R2_TNTI_PDF}. (b) Variation of the product $\left(L_R/h\right)\mathrm{d}\delta_\omega(x)/\mathrm{d}x$ with $\Rey_\theta$. Symbols: $\triangledown$ R2 ramp; $\color[rgb]{0.5,0.5,0.5}{\blacktriangledown}$ GDR ramp; \protect\dashedrule $\gamma_S = 1.1$. Fine, solid lines indicate the $\pm 0.07$ tolerance on $\gamma_S$.}
\label{fig:shearLayerProps}
\end{figure}

Now that we have identified a clear boundary layer footprint, it is time to get some insight into its effects on the flow after separation.  
In this respect, it seems interesting to begin by investigating the separated shear layer, because it is one of the main features of the flow. Following \citet{dandois07}, in a mean bidimensional flow the growth of the separated shear layer can be simply assessed by considering the streamwise evolution of the vorticity thickness $\delta_\omega$, defined as:
\begin{equation}
	\delta_\omega(x) = \frac{U_\infty(x)-U_{min}(x)}{\left(\partial U(x,y)/\partial y\right)_{max}},
	\label{eq:deltaOm}
\end{equation}
where $U_\infty$ and $U_{min}$ are the local maximum and the local minimum streamwise velocities. It is $U_{min}(x) <0 $ for $x < L_R$ and $U_{min}(x) \sim U_\infty$ in the entire separated flow (see \citet{le1997} and \citet{dandois07}).
Figure \ref{fig:shearLayerProps} (a) compares the streamwise evolution of $\delta_\omega/h$ observed on the two ramps. The streamwise coordinate is once again $x^*$. It appears that the evolutions of $\delta_\omega(x)/h$ collapse on two quasi-parallel, piecewise linear trends (also see \citet{dandois07}). The linear growth of the separated shear layer is consistent with the free shear layer approximation, in particular in a large neighbourhood of separation. We indicate with $\hat{x}^*$ the streamwise position at which the $\delta_\omega/h$ trend changes its slope.  The value of $\hat{x}^*$ appears to be at most a weak function of $\delta_e/h$: based on available measurements, we put $\hat{x}^* = 0.55 \pm 0.05$. For convenience, we also consider that $\hat{x}^*$ divides the flow domain in a separation region (for $x^*<\hat{x}^*$) and a reattachment region (for $x^*>\hat{x}^*$). In what follows, we will focus on the separation region only.

\subsection{Two competing length scales}\label{sec:twoLenghtScales}
The slopes of the $\delta_\omega/h$ trends shown in figure \ref{fig:shearLayerProps} (a) correspond to the non-dimensional growth rates of the separated shear layer. On $x^*<\hat{x}^*$, let us indicate this quantity with the symbol $\gamma_S$. It is:
\begin{equation}
\frac{\mathrm{d}\delta_\omega}{\mathrm{d}x}  = \gamma_S \left(L_R/h\right)^{-1}.
\label{eq:gamma1_b}
\end{equation}
Figure \ref{fig:shearLayerProps} (a) suggests that $\gamma_S$ is relatively insensitive to $\Rey_\theta$ effects, and at most a weak function of $\delta_e/h$. This is confirmed by direct measurements, reported in figure \ref{fig:shearLayerProps} (b), which give $\gamma_S = 1.1\pm 0.07$ across experiments.
If $\gamma_S$ can be considered almost constant, then the main difference between the R2 ramp and the GDR ramp on $x^*<\hat{x}^*$ is the intercept $\delta_\omega\left(0\right)/h$. 
In this regard, it is empirically found that $\delta_\omega\left(0\right)_{R2} \approx \delta_\omega\left(0\right)_{GDR} \approx$ \SI{6}{\milli\meter}, which immediately suggests that $\delta_\omega(0)$ does not scale with the height of the ramp $h$. 
It suits our purposes to consider that $\delta_\omega\left(0\right) = \beta\delta_e$, in which $\beta$ is a proportionality factor. In these experiments, it is $\beta  \approx 0.21$. The value of $\beta$ might depend on the maximum velocity gradient near the wall, and hence on the \textit{fullness} of the boundary layer velocity profile. Based on these considerations, on $x^*<\hat{x}^*$ it does not seem unreasonable to put:
\begin{equation}
	\frac{\delta_\omega(x)}{h} \approx \gamma_S x^* + \beta\frac{\delta_e}{h}.
	\label{eq:deltaOm_linear}
\end{equation}
Eq. \ref{eq:deltaOm_linear} has two important implications. Firstly, it appears that $\delta_\omega$ depends on two characteristic length scales, which can be intuitively related to different, coexisting phenomena: the $h$-scaled, linear term relates to the growth of a free-like shear layer after separation (the $h$ scaling, however, is specific to separating/reattaching flows), while the constant $\delta_e$ term shows the persisting influence of the incoming boundary layer dowstream of separation.
Secondly, the relative weight of these two terms appears to vary in the streamwise direction. In this respect, it is interesting to recast eq. \ref{eq:deltaOm_linear} as:
\begin{equation}
	\frac{\delta_\omega(x)}{h} \approx \gamma_S x^*\left(1 + \frac{\beta}{\gamma_S x^*}\frac{\delta_e}{h}\right).
	\label{eq:deltaOm_linear_2}
\end{equation}
Eq. \ref{eq:deltaOm_linear_2} suggests that the main contribution to $\delta_\omega$ is provided by $\beta\delta_e$ on a subdomain $x^*<\left[\left(\beta/\gamma_S\right)\left(\delta_e/h\right)\right]$, the extent of which, indicated with $x_\delta^*$, increases with $\delta_e/h$.
For $\delta_e/h = 0$, it is also $x_\delta^* = 0$ and the growth of $\delta_\omega$ after separation is correctly captured by the free shear layer analogy.
For $\delta_e/h \gg \gamma_S/\beta$, instead, the influence of the boundary layer should cover the entire recirculation region. In this case, it is thought that the flow might be better approximated by a boundary layer on a rough wall, rather than by a free shear layer.
Then, for asymptotic values of $\delta_e/h$ the flow has only one characteristic length scale, either $h$ or $\delta_e$.
For intermediate values of $\delta_e/h$, anyway, the two length scales are in competition: then, the flow in the separation region might evolve from a pure $h$ scaling, to a mixed scaling based on $h$ and $\delta_e$, to a scaling dominated by $\delta_e$, as $\delta_e/h$ increases.
It is pointed out that good collapse of $\delta_\omega/h$ on the entire recirculation region (figure \ref{fig:shearLayerProps}) suggests that an expression containing a similar dependency on both $h$ and $\delta_e$ might also exist for $x^*>\hat{x}^*$, so that present considerations might be extended, at least to a degree, to the reattachment region. 
It is stressed that these results are in good agreement with findings reported at § \ref{sec:competition}: $\delta_e/h$ provides a measure of the competition between the influence of the incoming boundary layer and the development of a free-like shear layer originating from the upper edge of the ramp. According to this idea, the higher is $\delta_e/h$, the further downstream the influence of the boundary layer persists after separation. In this instance, we report that available data give $x_\delta^* < 0.06$ on the GDR ramp, and $x_\delta^* < 0.16$ on the R2 ramp. Interestingly, it is $L_G^*/x_\delta^* \approx 5$ for both experiments, and hence $L_G^* \sim x_\delta^*$. Then, it seems possible to correlate the region in which the term $\beta\delta_e$ dominates eq. \ref{eq:deltaOm_linear_2} to a strong boundary layer footprint. 

\subsection{Effects on mean shear}\label{sec:effMeanShear}
The multi-scale dependency of $\delta_\omega$ expressed in eq. \ref{eq:deltaOm_linear_2} has direct consequences on the velocity gradient $\partial U/\partial y$, which in the bidimensional flow under investigation provides the main component of mean shear. By recasting eq. \ref{eq:deltaOm} and making use of eq. \ref{eq:deltaOm_linear_2}, one obtains the following expression:
\begin{equation}
	\left.\frac{\partial U(x,y)}{\partial y}\right|_{max} = \frac{U_\infty(x)-U_{min}(x)}{\delta_\omega} \approx \frac{U_\infty}{\gamma_S h x^*}\left(1+\frac{\beta}{\gamma_S x^*}\frac{\delta_e}{h}\right)^{-1}.
	\label{eq:meanShear_deltaOm}
\end{equation}
According to eq. \ref{eq:meanShear_deltaOm}, the presence of an incoming boundary layer increases the characteristic length scale of $\partial U/\partial y$, without fundamentally changing its velocity scale. Then, the higher is $\delta_e/h$, the less intensely the flow in the separation region is sheared.
Due to the central role of $\partial U/\partial y$, this suggests that a variation of $\delta_e/h$ will have far reaching consequences on the separated flow. In particular, lower mean shear is likely to induce lower turbulent production, and hence less intense Reynolds stresses in the whole separated flow. 
Interestingly, \citet{adamsJohnstonPart1} and \citet{stella2017} show that turbulent shear provides the strongest contribution to the pressure gradient at reattachment, so that the reduction of $\partial U/\partial y$ due to $\delta_e/h$ might indeed be at the origin of the progressive decrease of reattachment pressure observed by those authors (see also § \ref{sec:intro}). This topic, clearly connected to the matter of this paper, is currently being investigated. 

\section{Shear layer growth and mass entrainment}\label{sec:massEntrShLayer}
The previous section highlighted some important effects of $\delta_e/h > 0$ on the flow in the separation region. Anyway, figure \ref{fig:shearLayerProps} and eq. \ref{eq:deltaOm_linear_2} show that the growth of the separated shear layer is approximately linear, regardless to its strenght relative to the boundary layer footprint. If this is so, the free-shear layer analogy proposed by several researchers appears to still hold. It seems then possible to rely on one of the cornerstones of this analogy, that is that the growth rate $\mathrm{d}\delta_\omega/\mathrm{d}x$ depends on how mass in entrained into the shear layer. According to \citet{popeTurbulentFlows}, one can put:
\begin{equation}
	\frac{\mathrm{d}\delta_\omega}{\mathrm{d}x} \sim \sum v_E^*,
	\label{eq:SLgrowth_entr}
\end{equation}
were $\sum v_E^*$ is the total mean mass entrainment rate through the boundaries of the separated shear layer. For each of those boundaries, $v_E^* = v_E/U_\infty$ is a mean, large-scale mass entrainment rate.

Implications of eq. \ref{eq:SLgrowth_entr} need to be assessed in the light of findings at § \ref{sec:sepShLayAnalysis}. Indeed, while $\mathrm{d}\delta_\omega/\mathrm{d}x$ appears to be almost insensitive to $\delta_e/h$, mass entrainment, driven by the velocity field, is certainly affected by changes in mean shear.
Then, it does not seem unreasonable to expect that the proportionality factor between the two terms of eq. \ref{eq:SLgrowth_entr} will be a function of $\delta_e/h$.

In this section we aim at completing our investigation of the role of $\delta_e/h$ by shedding some light into this matter.
To this end, it seems convenient to start by computing the mean mass balance over the shear layer, to provide a global characterisation of mass exchanges. Then, we will analyse local mass fluxes, which lead to the computation of mass entrainment rates. 

\subsection{Mean mass balance}\label{sec:meanMassBalance}
\begin{figure}
\setlength{\unitlength}{1cm}
\begin{center} 
\includegraphics[width=\textwidth]{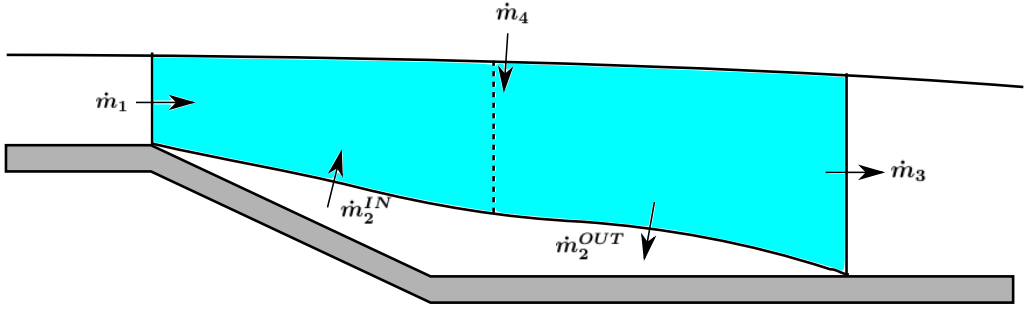} 
\end{center}
\caption{Control volume including the separated shear layer at $\Rey_\theta =$ \num{3617}.}
\label{fig:controlVolume}
\end{figure}

A first necessary step toward the computation of the mean mass balance consists in defining a control volume representative of the shear layer.
According to \citet{stella2017}, one valid choice is a volume $\mathcal{V}_c$ delimited by two vertical segments, placed at the positions of the mean separation point (called \textit{inlet}) and mean reattachment point (called \textit{outlet}); by the mean TNTI, which separates the shear layer from the free flow; and by the mean RRI.
For the sake of example, the control volume $\mathcal{V}_c$ for the GDR ramp flow at $\Rey_\theta =$ \num{3617} is shown in figure \ref{fig:controlVolume}.
Considering that the mean field is bidimensional, the total mass flux per spanwise unit length through each of the sides of $V_c$ is given by:
\begin{equation}
	\dot{m}_j = -\rho\int_{S_j} \! U_i(s) n_i(s) \mathrm{d}s = -\rho\int_{S_j} \! (U(s)\sin(\phi(s)) + V(s)\cos(\phi(s)))\, \mathrm{d}s,
	\label{eq:flux}
\end{equation}
where $S_j$ is the length of one side, $s$ a curvilinear abscissa, $\boldsymbol{n}(s)$ is the local normal to $S_j$ (pointing outward of $\mathcal{V}_c$) and $\phi$ is the angle between $\boldsymbol{n}(s)$ and the Y axis. The index $j$ goes from \num{1} to \num{4}. $j=1$ indicates the inlet at the mean separation point; then, $j$ increases counterclockwise, so that $j=4$ identifies the mean TNTI. Of course, continuity implies that $\sum m_j = 0$.

Measured mean mass fluxes are reported in table \ref{tab:massFluxComp}.
Uncertainties on mass balance are mainly caused by corrupted velocity vectors produced by laser reflections on the wall, in particular in a neighbourhood of the mean separation point. 
As already pointed out in \citet{stella2017}, $\dot{m}_2$ must be zero, because in average the backflow balances shear layer entrainment through the RRI in a neighbourhood of the mean separation point (see \citet{chapman1958} and \citet{adamsJohnstonPart2}). 
Our results also agree with \citet{stella2017} in evidencing that $\dot{m}_4$ is not negligible.
Indeed, since the TNTI is not a streamline, mass entrainment through the TNTI compensates the difference of mass fluxes between the outlet and the inlet, i.e. $\dot{m}_4 = - \dot{m}_3 - \dot{m}_1 \neq 0$. 
The role of the TNTI appears to be even stronger on the GDR ramp than on the R2 ramp: while on the latter the TNTI contributes to mass balance with approximately \SI{30}{\percent} of the mass injected into $\mathcal{V}_c$ by the incoming boundary layer, on the former $\dot{m}_4$ becomes the dominant positive mass contribution.
\begin{table}
\centering
\begin{tabu}{@{}crrrrrrrrrrr@{}}
&& \multicolumn{5}{c}{R2} && \multicolumn{4}{c}{GDR}\\
\midrule
$\Rey_\theta$ && \num{2006} & \num{3262} & \num{4122} & \num{4738} & \num{5512} && \num{1788} & \num{2547} & \num{3617} & \num{4340} \\
\midrule
$\dot{m}_1^*$ &&	 0.39 &  0.43 &  0.41 &  0.45 &  0.44 &&  0.17 &  0.15 &  0.15 &  0.14\\ 	  
$\dot{m}_3^*$ &&	-0.53 & -0.56 & -0.55 & -0.64 & -0.59 && -0.45 & -0.47 & -0.45 & -0.49\\
$\dot{m}_4^*$ && 	 0.14 &  0.14 &  0.14 &  0.20 &  0.17 &&  0.27 &  0.28 &  0.27 &  0.33\\
\midrule
$\epsilon_m^* = (\dot{m}_1^* + \dot{m}_3^* + \dot{m}_4^*)$ && 0.00 & 0.00 & 0.007 & 0.006 & 0.019 && -0.03 & -0.04 & -0.03 & -0.02\\
&& & & & & && & & &\\
$\dot{m}_2^*$ &&  -0.005 &  -0.007 &  -0.007 &  -0.004 &  0.005 && -0.01 & -0.005 & 0.008 & 0.014\\
\end{tabu}
\caption{Mass fluxes normalized on $\rho h U_\infty$. The error $\epsilon_m^*$ does not include $\dot{m}_2^*$ because $\dot{m}_4^*$ should balance the difference between the inlet and the outlet, independently of how well the $\dot{m}_2^* = 0$ condition is met.}
\label{tab:massFluxComp}
\end{table}

The decreasing weight of the incoming boundary layer on the global mass balance seems consistent with a lower value of the parameter $\delta_e/h$, as follows. 
It appears from table \ref{tab:massFluxComp} that the mass flow at the inlet $\dot{m}_1$ does not scale with $\rho h U_\infty $. Of course, a straightforward alternative is scaling $\dot{m}_1$ on $\rho \delta_e U_\infty$, as reported in table \ref{tab:massInlet_delta}. This second normalisation appears to be more relevant: it is $\dot{m}_1/\rho \delta_e U_\infty \approx 0.5 \pm 0.05$. The value of this ratio is of the same order of magnitude of $Y_{T,e}/\delta_e$. Although scatter is not always negligible, it seems then possible to assume that $\dot{m}_1$ is sized by $Y_{T,e}$, and more in general by the outer scales of the incoming boundary layer.
\begin{table}
\centering
\begin{tabu}{@{}crrrrrrrrrrr@{}}
&& \multicolumn{5}{c}{R2} && \multicolumn{4}{c}{GDR}\\
\midrule
$\Rey_\theta$ && \num{2006} & \num{3262} & \num{4122} & \num{4738} & \num{5512} && \num{1788} & \num{2547} & \num{3617} & \num{4340} \\
\midrule
$\dot{m}_1/\rho \delta_e U_\infty$ &&	 0.45  & 0.50 &  0.50 &  0.55 &  0.48 &&  0.48 &  0.56 &  0.45 &  0.54\\
\end{tabu}
\caption{Inlet mass fluxes ($\dot{m}_1$) normalized on $\rho \delta_e U_\infty$.}
\label{tab:massInlet_delta}
\end{table} 
Then, the amount of mass transported by the incoming boundary layer into $\mathcal{V}_c$ is comparable in the two experiments, because $\delta_e$ is also approximately the same. However, table \ref{tab:massFluxComp} also suggests that more mass leaves from the outlet as $h$ increases, that is, at least in our experiments, for decreasing values of $\delta_e/h$. Since the mean separated shear layer is stationary, this implies that, to verify continuity, the mass contribution of the TNTI must also increase as $\delta_e/h$ decreases.

\subsection{Local mean mass fluxes}\label{sec:locMeanFluxes}
Now that some of the effects of $\delta_e/h$ on the mean mass balance have been identified, it is convenient to consider the local mass fluxes along the RRI and the TNTI. 
The normalised local flux at any point of the two interfaces can be estimated as:
\begin{equation}
\dot{m}_{xi}^* = \frac{1}{\rho U_\infty}\frac{\mathrm{d}\dot{m}_i}{\mathrm{d}s} = -\frac{1}{U_\infty}(U(s)\sin(\phi(s)) + V(s)\cos(\phi(s)).
\label{eq:locFlux}
\end{equation}
This finer analysis can provide information which is hidden by the integral approach of § \ref{sec:meanMassBalance}, such as the spatial distribution of mass fluxes and their scaling laws.

\subsubsection{RRI fluxes}\label{sec:RRIFlux}
\begin{figure}
\setlength{\unitlength}{1cm}
\begin{center} 
\includegraphics[width=\textwidth]{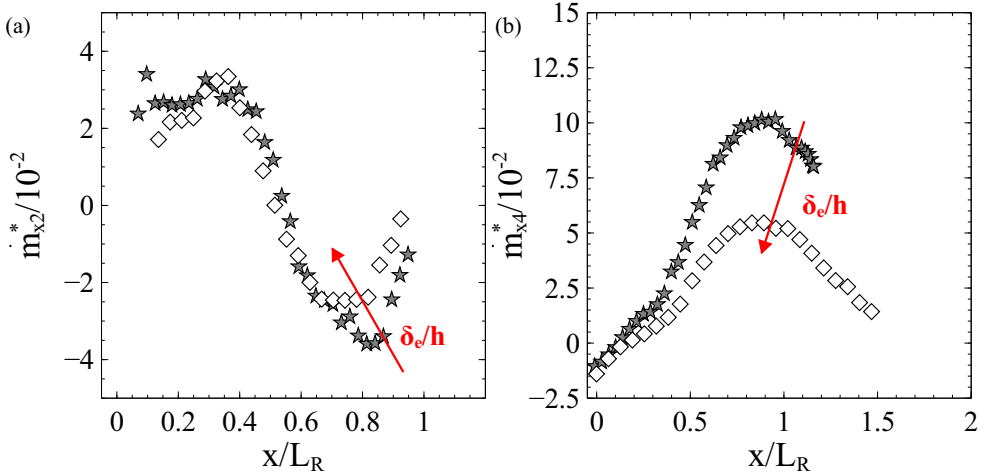} 
\end{center}
\caption[Local mass fluxes]{Normalized local mass fluxes along the mean interfaces of the separated shear layer at for $\Rey_\theta = $ \num{3262} (R2 ramp - \graylozenge) and $\Rey_\theta = $ \num{3617} (GDR ramp - \graystar). (a) $\dot{m}_{x2}^*$ along the RRI. Only one in each six points is represented in the case of the R2 ramp, and only one in fifteen in the case of the GDR ramp. (b) $\dot{m}_{x4}^*$ along the TNTI. Only one in each ten points is represented in the case of the R2 ramp, and only one in twenty in the case of the GDR ramp. Red arrows indicate the effect of increasing $\delta_e/h$.}
\label{fig:localFluxes}
\end{figure}

The streamwise evolution of $\dot{m}_{x2}^*$ is reported in figure \ref{fig:localFluxes}(a) for $\Rey_\theta = $ \num{3262} (R2 ramp) and $\Rey_\theta = $ \num{3617} (GDR ramp). The distribution of $\dot{m}_{x2}^*$ along the mean RRI is approximately odd, with a change of sign at $x^* \approx \hat{x}^*$. As such, $\dot{m}_{x2}^*$ appears to be very well correlated to the development of the separated shear layer (see figure \ref{fig:shearLayerProps} (a)).
Curves from both experiments collapse together nicely, with the exception of the domain $x^* > 0.7$. 
Since mass entrainment at reattachment appears to be correlated to turbulent shear (\citet{stella2017}), this difference might be a consequence of the different value of $\delta_e/h$ (also see pressure distributions and turbulent shear stress profiles in \citet{adamsJohnstonPart1}).

Figure \ref{fig:localFluxes} (a) proves that, even if $\dot{m}_{2}^* = 0$, local mass transfer through the RRI is not negligible. On $x^* < \hat{x}^*$, $\dot{m}_{x2}^*$ injects mass into $\mathcal{V}_c$: we will indicate quantities relative to this domain with the symbol $^{IN}$. On the contrary, $\dot{m}_{x2}^*$ extracts mass from $\mathcal{V}_c$ on $x^* > \hat{x}^*$: quantities relative to this domain will be marked by the symbol $^{OUT}$. Considering the scaling of figure \ref{fig:localFluxes}(a), the entrainment rate on, say, $x^* < \hat{x}^*$ is simply given by:
\begin{equation}
{v_{E,R}^{*}}^{IN} = -\frac{1}{\rho U_\infty \hat{S}}\int_{0}^{\hat{s}} \! \rho U_i(s) n_i(s) \mathrm{d}s,
\label{eq:Ve_A_R}
\end{equation}
where $\hat{s}$ is the curvilinear abscissa at $x^* = \hat{x}^*$ and:
\begin{equation}
\hat{S} = \int_{0}^{\hat{s}}\mathrm{d}s.
\end{equation}
${v_{E,R}^{*}}^{OUT}$ can be computed with an expression similar to eq. \ref{eq:Ve_A_R}. Available data give ${v_{E,R}^{*}}^{IN} \approx 0.0223 \pm 0.002$ and ${v_{E,R}^{*}}^{OUT} \approx -0.0225 \pm 0.0015$ on the GDR ramp; ${v_{E,R}^{*}}^{IN} \approx 0.0224 \pm 0.002$ and ${v_{E,R}^{*}}^{OUT} \approx -0.0205 \pm 0.0015$ on the R2 ramp. These measurements clearly indicate that the mean entrainment rate through the RRI is independent of $\delta_e/h$, in spite of the influence of this latter parameter at reattachment. In general, $v_{E,R}^*$ is also weakly affected by other parameters such as $h$, $\Rey_h$ (at least if $\Rey_h>36000$, see \citet{nadge14}), the value of $\Rey_\theta$ in the incoming boundary layer (see table \ref{tab:veT}) and, to a large extent, $L_R$ (see \citet{stella2018}). 

\subsubsection{TNTI fluxes}\label{sec:TNTIflux}
Figure \ref{fig:localFluxes}(b) presents the streamwise evolution of $\dot{m}_{x4}^*$, obtained by applying eq. \ref{eq:locFlux} to the TNTI. As in the case of the RRI, the behaviour of $\dot{m}_{x4}^*$ changes at $x^* \approx \hat{x}^*$, with sizeably more intense transfer in the reattachment region. In this respect, comparison with \citet{stella2017} suggests that the peak of mass entrainment through the TNTI is reached in proximity of the position of maximum pressure gradient.
Once again, the two ramps show different trends in this region, which might be related to $\delta_e/h$.
It is evident that the intensity of local fluxes is higher in the case of the GDR ramp. This seems consistent with the increased contribution brought by the TNTI to mass balance, as $\delta_e/h$ decreases (see § \ref{sec:meanMassBalance}). 
Anyway, the reattachment region accounts for \SIrange{80}{90}{\percent} of $\dot{m}_4$ on the R2 ramp, but for only \SIrange{60}{80}{\percent} of $\dot{m}_4$ on the GDR ramp. Then, mass entrainment through the TNTI is fundamentally concentrated in the reattachment region in the case of the R2 ramp, while it acts more homogenously over the GDR ramp. 
The mean entrainment rate on $x^* \in \left(0, \hat{x}^*\right)$, indicated with $v_{E,T}^{*}$, can be computed simply, by adapting eq. \ref{eq:Ve_A_R} to the TNTI. Table \ref{tab:veT} shows that, unlike ${v_{E,R}^*}^{IN}$, $v_{E,T}^{*}$ increases with $\Rey_\theta$. In addition, the two ramps differ by the relative weight of $v_{E,T}^*$ and ${v_{E,R}^*}^{IN}$. Generally speaking, $v_{E,T}^*$ is of the same order of magnitude as ${v_{E,R}^*}^{IN}$ on the GDR ramp, but it is sizeably lower on the R2 ramp. This might appear counterintuitive, at first sight, as one could expect $v_{E,T}^{*}$ to rise accordingly to the strenght of boundary layer footprint. Anyway, it should be reminded that $\partial U/\partial y$ decreases as $\delta_e/h$ increases. This tends to hinder turbulent production, which reduces mixing and hence the intensity of transfer among different regions of the flow.
\begin{table}
\centering
\begin{tabular}{@{}cccccccccccc@{}}
&& \multicolumn{5}{c}{R2} && \multicolumn{4}{c}{GDR}\\
\midrule
$\Rey_\theta$ && \num{2006} & \num{3262} & \num{4122} & \num{4738} & \num{5512} && \num{1788} & \num{2547} & \num{3617} & \num{4340} \\      
\midrule
$v_{E,T}^{*}/10^{-2}$ && 0.62 & 0.86 & 1.08 & 1.57 & 1.85 && 1.79 & 1.94 & 1.88 & 2.62 \\
${v_{E,R}^{*}}^{IN}/10^{-2}$ && 2.26 & 2.24 & 2.17 & 2.43 & 2.11 && 2.07 & 2.22 & 2.40 & 2.23\\
\end{tabular}
\caption[TNTI entrainment rates]{Mean entrainment rates at the mean TNTI and at the mean RRI on $x^* \in \left(0, \hat{x}^*\right)$.}
\label{tab:veT}
\end{table}

\subsection{Total entrainment rates and shear layer growth}\label{sec:totEntrRate}
\begin{figure}
\setlength{\unitlength}{1cm}
\begin{center} 
\includegraphics[width=0.6\textwidth]{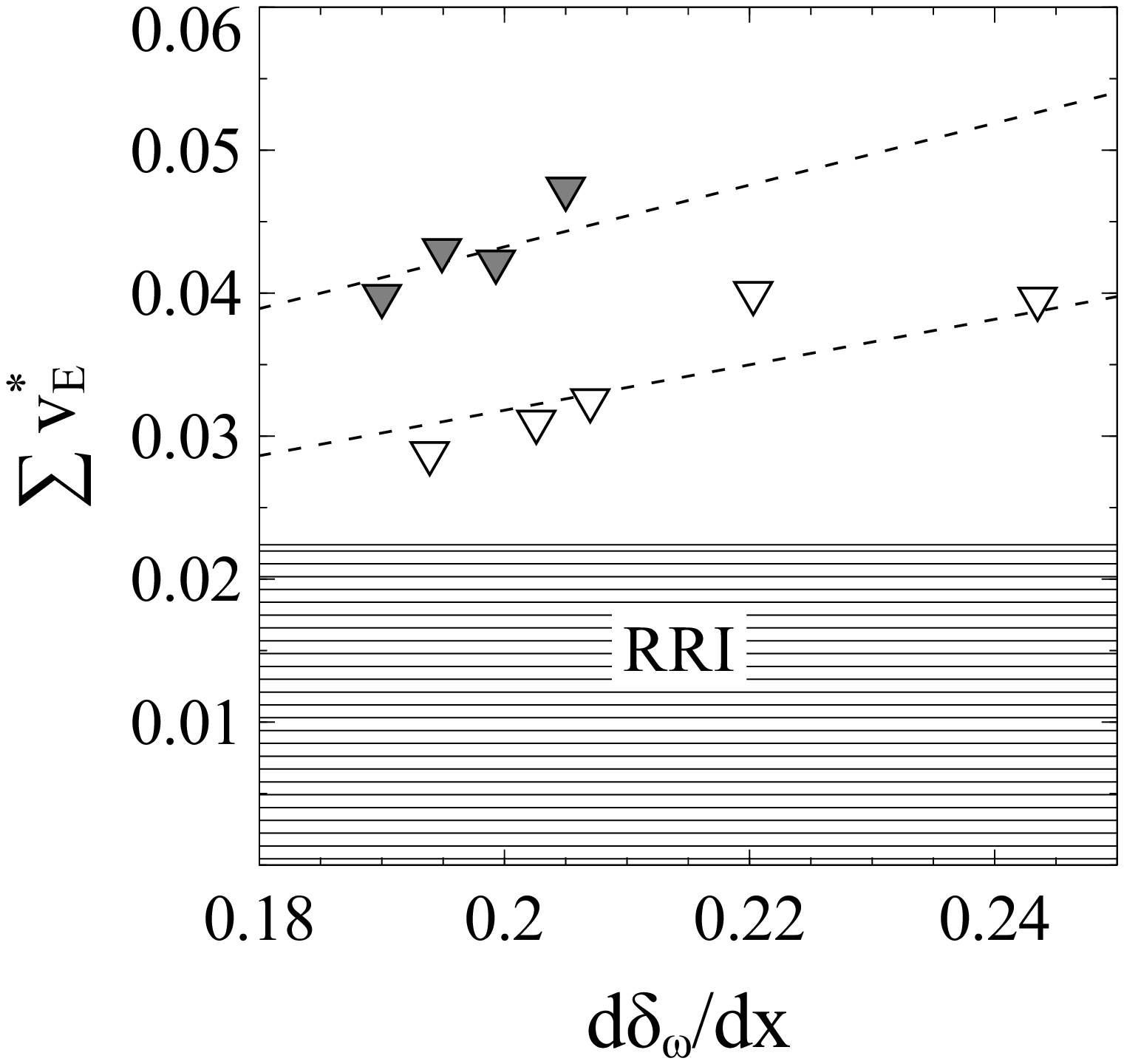} 
\end{center}
\caption{Correlation between $\mathrm{d}\delta_\omega/\mathrm{d}x$ and $\sum v_E^*$ (eq. \ref{eq:SLgrowth_entr}). Symbols: $\triangledown$ R2 ramp; $\color[rgb]{0.5,0.5,0.5}{\blacktriangledown}$ GDR ramp; {\protect\dashedrule} linear best fits in the form $y = 1/K_\omega \,x$, with $1/K_\omega$ being the slope. The hatched area in (a) indicates ${v_{E,R}^*}^{IN}$.}
\label{fig:dDelta_dx_ve}
\end{figure}

\begin{figure}
\setlength{\unitlength}{1cm}
\begin{center} 
\fbox{\includegraphics[width=0.98\textwidth]{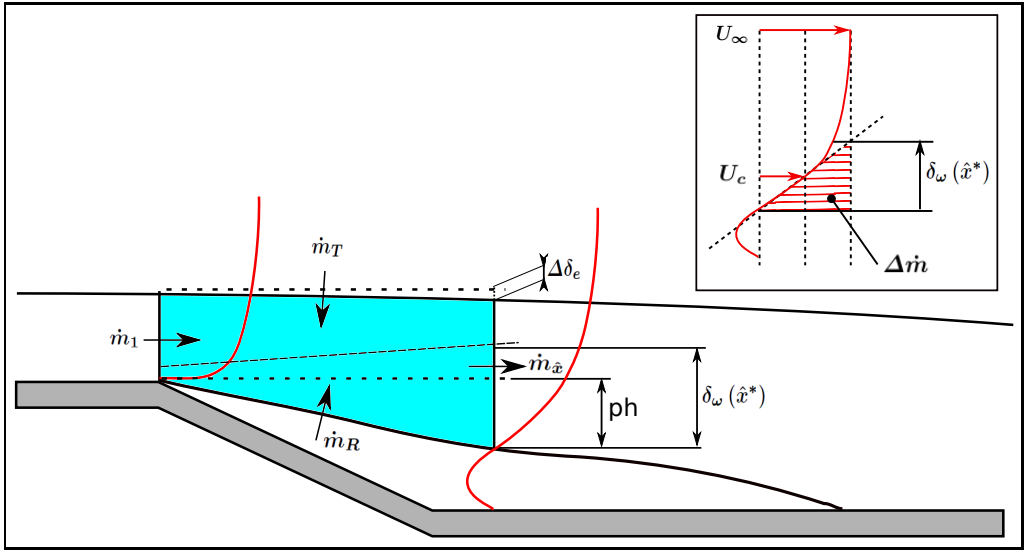}}
\end{center}
\caption{View of the control volume $\mathcal{V}_{C2}$, that covers the separation region. $\dot{m}_{1}$, $\dot{m}_R$, $\dot{m}_T$ and $\dot{m}_{\hat{x}}$ are the mass fluxes through the inlet, the RRI, the TNTI and the outlet, respectively. The RRI is adopted as lower boundary of shear layer development. The upper boundary is indicated with a dashed line. Red lines are used to represent streamwise velocity profiles at the inlet and at the outlet. The insert details the definitions of $U_c$ and $\Delta\dot{m}$.}
\label{fig:controlVol2_sketch}
\end{figure}

\begin{figure}
\setlength{\unitlength}{1cm}
\begin{center} 
\includegraphics[width=0.6\textwidth]{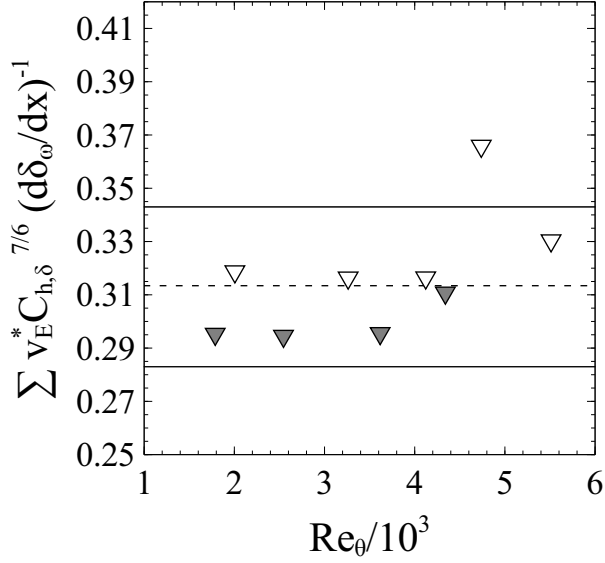} 
\end{center}
\caption{Evolution of the ratio $\sum v_E^* {C_{h,\delta}}^{7/6} \left(\mathrm{d}\delta_\omega/\mathrm{d}x\right)^{-1}$ (eq. \ref{eq:outletMass_7}) with $\Rey_\theta$. Symbols: $\triangledown$ R2 ramp; $\color[rgb]{0.5,0.5,0.5}{\blacktriangledown}$ GDR ramp; {\protect\dashedrule} best fits in the form $y = 1/K_\omega$. Fine, solid lines indicate the $\pm 0.03$ tolerance on $1/K_\omega$. }
\label{fig:dDelta_dx_ve_corr}
\end{figure}

In the previous paragraphs we showed that the parameter $\delta_e/h$ sizeably affects mass entrainment from the free-stream, while leaving entrainment from the reverse flow unchanged. 
Let us now get back to the correlation between $\mathrm{d}\delta_\omega/\mathrm{d}x$ and $\sum v_E^* = {v_{E,R}^*}^{IN} + v_{E,T}^*$. As expected, figure \ref{fig:dDelta_dx_ve} shows that eq. \ref{eq:SLgrowth_entr} is well verified in both experiments.
Findings reported at § \ref{sec:RRIFlux} suggest that the variation of $\sum v_E^*$ is fundamentally driven by $v_{E,T}^*$.
Considering that $\gamma_S$ is approximately constant and that $L_R/h$ is a function of $\Rey_\theta$ (see figure \ref{fig:LR_RE_THETA} (b)), eq. \ref{eq:gamma1_b} clearly indicates that $\Rey_\theta$ determines, for each ramp, the position of available datapoints along the linear trends of figure \ref{fig:dDelta_dx_ve}.
This confirms that, at least if other parameters such as the turbulent state of the incoming boundary layer or the geometry of the ramp are kept constant, the proportionality factor between the two terms of eq. \ref{eq:SLgrowth_entr} (i.e. the slopes of the linear trends of figure \ref{fig:dDelta_dx_ve}) is mainly affected by $\delta_e/h$.
Our next objective is to investigate whether such trends can be collapsed together by a scaling factor that takes into account the effects of $\delta_e/h$.
Since it was observed that $\delta_e/h$ changes mass entrainment from the free stream, a mean mass balance seems a promising starting point for our discussion.
Unlike at § \ref{sec:meanMassBalance}, we now focus on the separation region exclusively, i.e. on $x^* < \hat{x}^*$. With reference to figure \ref{fig:controlVol2_sketch}, let us define a new control volume $\mathcal{V}_{C2}$, that corresponds to the portion of $\mathcal{V}_C$ for $x^* < \hat{x}^*$. To begin with, we can directly write:
\begin{equation}
\dot{m}_T + \dot{m}_R = \dot{m}_{\hat{x}}-\dot{m}_{1},
\label{eq:massBalanceForChd}
\end{equation}
where $\dot{m}_T$, $\dot{m}_R$ and $\dot{m}_{\hat{x}}$ are the norms of mass fluxes through the TNTI, the RRI and the outlet of $\mathcal{V}_{C2}$, respectively.
According to table \ref{tab:massInlet_delta}, it is straightforward to put:
\begin{equation}
\dot{m}_{1} \approx \rho U_\infty \frac{\delta_e}{2}.
\label{eq:inletMass}
\end{equation}
In principle, computing $\dot{m}_{\hat{x}}$ requires to know the shear layer velocity profile at $x^* = \hat{x}^*$, indicated with $U\left(\hat{x}^*,s\right)$. A tempting starting point to estimate $U\left(\hat{x}^*,s\right)$ is the velocity profile of the free shear layer, which is usually approximated by an error function. For small values of $\delta_e/h$ and low turbulent intensities, \citet{tanner1973} (among others) shows that this profile can be adapted to massively separated flows and provide good predictions of mean flow properties, such as reattachment wall-pressure. However, these hypotheses are not generally acceptable in the present framework, as they defeat the very purpose of our discussion.
An alternative approach to estimate $\dot{m}_{\hat{x}}$ might be based on general considerations on the topology of the flow.
To begin with, if $U\left(\hat{x}^*,s\right) = U_\infty$, the amount of mass entrained by the flow through the outlet section of $\mathcal{V}_{C2}$, indicated with $S_{\hat{x}^*}$, would be simply given by:
\begin{equation}
\dot{m}_{U_\infty} = \rho U_\infty S_{\hat{x}^*}.
\label{eq:mass_Uinf}
\end{equation}
In itself, $\dot{m}_{U_\infty}$ is not a good estimate of $\dot{m}_{\hat{x}}$, because $\partial U / \partial y$ is not negligible along a wide portion of $S_{\hat{x}^*}$. Then, it does not seem unreasonable to introduce a mass entrainment \textit{deficit}, representing the amount of mass that \textit{does not} cross $S_{\hat{x}^*}$ due to the velocity gradient. Eq. \ref{eq:inletMass} suggests that the velocity scale is approximately $U_\infty$ in the outer part of $S_{\hat{x}^*}$. Then, the vertical velocity gradient mainly depends on the separated shear layer and, based on findings at § \ref{sec:sepShLayAnalysis}, we can tentatively put:
\begin{equation}
\Delta\dot{m} \approx \rho \left(U_\infty - U_c\right) \delta_\omega\left(\hat{x}^*\right),
\end{equation}
in which $U_c$ is a characteristic convection velocity. Under the free-shear layer analogy, we can refer to \citet{popeTurbulentFlows} and define $U_c$ as:
\begin{equation}
U_c = \frac{U_\infty + U_{min}}{2}.
\label{eq:convVelShearLayer}
\end{equation}
This expression needs to be corrected, to take into account that the lower boundary of $\mathcal{V}_{C2}$ is placed at $U=0$, rather than $U = U_{min}$. It is then $U_c \approx U_\infty/2$ and hence:
\begin{equation}
\Delta\dot{m} \approx \rho \frac{U_\infty}{2}\delta_\omega\left(\hat{x}^*\right).
\end{equation}
With these results, we can propose the following formulation for $\dot{m}_{\hat{x}}$:
\begin{equation}
\dot{m}_{\hat{x}} \approx \dot{m}_{U_\infty} - \Delta\dot{m} \approx \rho U_\infty \left[S_{\hat{x}^*} - \frac{\delta_\omega\left(\hat{x}^*\right)}{2}\right].
\label{eq:outletMass_1}
\end{equation}
All terms in this expression are known, with the exception of the length of the outlet section. According to figure \ref{fig:controlVol2_sketch}, $S_{\hat{x}^*}$ can be estimated with three terms, as follows:
\begin{equation}
S_{\hat{x}^*} \approx \frac{\delta_e}{2} + ph - \Delta\delta_e.
\label{eq:outputSection}
\end{equation}
The half-thickness of the incoming boundary layer $\delta_e/2$ is linked to $Y_{T,e}$. The term $ph$, with $0 < p < 1$, takes into account the development of the separated shear layer toward the wall. Measurements on available velocity fields give $p \approx 0.45 \pm 0.05$. Finally, the term $\Delta\delta_e$ takes into account the inclination of the mean TNTI toward the wall. This term cannot be predicted simply, but for a first approximation we can use geometrical consideration on the shape of $\mathcal{V}_{C2}$ to put:
\begin{equation}
\Delta\delta_e/\delta_e \approx \Phi_{TNTI} \, \hat{x}^* \, L_R/h \, \left(\delta_e/h\right)^{-1},
\end{equation}
in which $\Phi_{TNTI}$ is the slope of the mean TNTI, approximated with a straight line over the extent of $\mathcal{V}_{C2}$. The investigation of the behaviour of $\Phi_{TNTI}$ is beyond the scope of this work, but preliminary measurements seem to show that:
\begin{equation}
\Phi_{TNTI} \, \left(L_R/h\right) \approx k_\Phi,
\end{equation}
where it is $k_\Phi \approx 0.10$ on the R2 ramp and $k_\Phi \approx 0.03$ on the GDR ramp. Surprisingly, the product $ k_\Phi \left(\delta_e/h\right)^{-1}$ is almost constant across experiments, so that $\Delta\delta_e/\delta_e$ varies within a small range approximated by $\left(0.0675 \pm 0.0025\right)$.
By injecting these results into eq. \ref{eq:outletMass_1} and by making use of eq. \ref{eq:deltaOm_linear}, one finds:
\begin{equation}
\dot{m}_{\hat{x}} \approx \rho U_\infty \left[\delta_e\left(\frac{1 - \beta}{2} -\frac{\Delta\delta_e}{\delta_e}\right) + h\,\left(p - \frac{\gamma_S \, \hat{x}^*}{2}\right)\right].
\label{eq:outletMass_2} 
\end{equation}
Let us now get back to eq. \ref{eq:massBalanceForChd}. By plugging in eq. \ref{eq:inletMass} and eq. \ref{eq:outletMass_2}, simple manipulations lead to:
\begin{equation}
\dot{m}_T + \dot{m}_R \approx \frac{\rho U_\infty}{2} \, h \gamma_S \left(\frac{2p}{\gamma_S} - \hat{x}^*\right) \left[1-\frac{\beta + 2\Delta\delta_e/\delta_e}{2p - \gamma_S\hat{x}^*} \, \frac{\delta_e}{h}\right],
\label{eq:outletMass_3}
\end{equation}
in which terms were reorganised as to make the dependency on $\delta_e/h$ explicit. We indicate with the symbols $S_T$ and $S_R$ the portions of the TNTI and of the RRI, respectively, that delimit $\mathcal{V}_{C2}$. If it assumed that $S_T \approx S_R \approx L_R/2$, it is:
\begin{equation}
2\frac{\dot{m}_T + \dot{m}_R}{\rho U_\infty L_R} \approx \sum v_E^*. 
\end{equation}
By making use of this result, eq. \ref{eq:outletMass_3} becomes:
\begin{equation}
\sum v_E^* \approx \frac{h}{L_R} \gamma_S \left(\frac{2p}{\gamma_S} - \hat{x}^*\right) \left[1-\frac{\beta + 2\Delta\delta_e/\delta_e}{2p - \gamma_S\hat{x}^*} \, \frac{\delta_e}{h}\right].
\label{eq:outletMass_4}
\end{equation}
According to eq. \ref{eq:gamma1_b}, it is $\mathrm{d}\delta_\omega/\mathrm{d}x = \gamma_S \left(L_R/h\right)^{-1}$. Then, it is possible to write:
\begin{equation}
\sum v_E^* \left[1-\frac{\beta + 2\Delta\delta_e/\delta_e}{2p - \gamma_S\hat{x}^*} \, \frac{\delta_e}{h}\right]^{-1} \approx \frac{\mathrm{d}\delta_\omega}{\mathrm{d}x} \left(\frac{2p}{\gamma_S} - \hat{x}^*\right).
\label{eq:outletMass_5}
\end{equation}
If $\delta_e/h$ is sufficiently small, a Taylor expansion can be used to rewrite eq. \ref{eq:outletMass_5} as:
\begin{equation}
\sum v_E^* {C_{h,\delta}}^{q_S} \approx \frac{1}{K_\omega} \, \frac{\mathrm{d}\delta_\omega}{\mathrm{d}x}.
\label{eq:outletMass_6}
\end{equation}
In this expression, it is $C_{h,\delta} = \left(1 + \delta_e/h\right)$, $ q_S = \left(\beta + 2\Delta\delta_e/\delta_e\right)/\left(2p - \gamma_S\hat{x}^*\right)$ and $1/K_\omega = \left(2p/\gamma_S - \hat{x}^*\right)$. Available data give $q_S \approx 1.17 \approx 7/6$ across experiments. The proportionality factor $1/K_\omega$ is independent of $\delta_e/h$, which suggests that $1/K_\omega$ might be assimilable to the proportionality factor typical of free shear layers. For a free shear layer with similar values of $U_\infty$ and $U_{min}$ \citet{popeTurbulentFlows} predicts:
\begin{equation}
\frac{1}{{K_{\omega,0}}} \approx 0.24\,\frac{U_\infty - U_{min}}{U_\infty} \approx 0.28,
\label{eq:K_omega_pred}
\end{equation}
where, following \citet{le1997} and \citet{dandois07}, it was considered $U_{min}/U_\infty \approx -0.15$. Significantly, available data give $1/K_\omega \approx 0.27$. This result agrees very well with the expected value for free shear layer $1/K_{\omega,0}$.
In this respect, it is interesting to obtain a second estimation of $1/{K_\omega}$, by recasting eq. \ref{eq:outletMass_6} as:
\begin{equation}
\frac{1}{K_\omega} \approx \sum v_E^* {C_{h,\delta}}^{q_S} \left(\frac{\mathrm{d}\delta_\omega}{\mathrm{d}x}\right)^{-1}.
\label{eq:outletMass_7}
\end{equation}
Figure \ref{fig:dDelta_dx_ve_corr} shows the evolution of eq. \ref{eq:outletMass_7} with $\Rey_\theta$. Most datapoints appear to be clustered within $1/K_\omega \approx 0.31 \pm 0.03$, which is once again in good accordance with $1/K_{\omega,0}$ and with the prediction $1/K_\omega \approx 0.27$ provided by eq. \ref{eq:outletMass_6}. 
These considerations suggest that eq. \ref{eq:outletMass_6} is quite effective at scaling shear layer behaviours across experiments. Then, it seems possible to put:
\begin{equation}
\sum v_E^* {C_{h,\delta}}^{7/6} \approx \frac{1}{K_{\omega,0}} \, \frac{\mathrm{d}\delta_\omega}{\mathrm{d}x}.
\label{eq:linModel_2}
\end{equation}
This expression predicts that, in the separation region, the effect of $\delta_e/h$ on rates of mass entrainment into the separated shear layer are scaled by a power law of the scaling factor $C_{h,\delta}$. Once such effects are compensated for, the relationship between mass entrainment and shear layer collapse on a trend typical of free shear layers.

\section{Conclusions}\label{sec:conclu}
In this work we experimentally investigated how the incoming boundary layer affects a massively separated turbulent flow. Since some of these effects have already been identified at reattachment, we focused on a large neighbourhood of separation.
The chosen study case was the massively separated turbulent flow generated by a sharp edge, \SI{25}{\degree} descending ramp. Significantly, this work could rely on two experimental models with sizeably different values of ramp height $h$ (their ratio was 1:3), but otherwise substantially similar geometries and incoming flows. This allowed us to compare the effects of two very different $\delta_e/h$ ratios, in which $\delta_e$ is the full thickness of the incoming boundary layer at separation.
 
By using the Turbulent/Non-Turbulent Interface (TNTI) as a tracer, we showed that the incoming boundary layer leaves a clear footprint on the separated flow. In particular, the statistical distribution of the TNTI keeps the gaussian form typical of boundary layers on an extent $L_G$ after separation. In both experiments, it is $L_G/L_R \approx \delta_e/h$. Since $L_R$ is the characteristic scale of shear layer development, the ratio $\delta_e/h$ seems to be representative of the relative strength of the incoming boundary layer with respect to the separated shear layer.

On these bases, we set out to understand how the footprint of the boundary layer affects the separated flow. In particular, we considered shear layer development, classically assessed with a vorticity thickness $\delta_\omega$. We found that shear layer growth remains linear and that $\mathrm{d}\delta_\omega/\mathrm{d}x$ appears to be at most a weak function of $\delta_e/h$, at least on the $\delta_e/h$ range covered in this study. Anyway, depending on the value of $\delta_e/h$, the separated flow might pass from a pure $h$ scaling, to a mixed $h$-$\delta_e$ scaling, to a $\delta_e$-dominated scaling. In this latter case, we argue that the flow might be better interpreted as a thick boundary layer on a rough wall. 
In addition, we used simple dimensional considerations to show that the higher is $\delta_e/h$, the less the mean separated flow is sheared. 

It is generally admitted that $\mathrm{d}\delta_\omega/\mathrm{d}x$ is proportional to the total mass entrainment rate toward the shear layer. For this reason, in the last part of this work we analysed how $\delta_e/h$ affects mass entrainment in the separation region. With a simple mass balance, we showed that the higher is $\delta_e/h$, the lower is the entrainment rate from the free flow, which seems consistent with a decrease of mean velocity gradients across different regions of the flow. Anyway, we demonstrated that a power law of the scaling factor $C_{h,\delta} = \left(1 + \delta_e/h\right)$ can be used to scale the effects of $\delta_e/h$ on mass entrainment rates. By doing so, the relationship between mass entrainment rates and shear layer growth appears to collapse on a trend typical of free shear layers.

All in all, this work suggests that separating/reattaching flows assimilable to the one under study might be thought of as the result of the competition of at least two simpler flows: a free-like separated shear layer, scaled by $h$, and the incoming boundary layer, scaled by $\delta_e$, their equilibrium being determined by $\delta_e/h$. At least to a certain extent, it appears possible to reduce these non-trivial, multi-scale flows to one or the other of their canonical components, by introducing simple scaling factors based on $\delta_e/h$, such as $C_{h,\delta}$.
These findings might indicate that the optimal solutions for controlling or predicting such separating/reattaching flows might strongly depend on the parameter $\delta_e/h$. In future works, we will try to confirm this view with wider parametric studies, possibly based on numerical simulations. In particular, it seems important to span a wider $\Rey_\theta$ range, as well as to test our results against variations of geometrical parameters such as ramp profile and $ER$.   

\section*{Acknowledgments}
This work was supported by the French National Research Agency (ANR) through the \textit{Investissements d’Avenir} program under the Labex CAPRYSSES Project (ANR-11-LABX-0006-01), and by the CNRS through the collaborative project \textit{Groupement de Recherche 2502}.

\appendix

\bibliographystyle{jfm}
\bibliography{FS_thesis_biblio}
\clearpage
 
\end{document}